\documentclass[prd,aps,preprint,tightenlines,groupedaddress,nofootinbib,showpacs,floatfix]{revtex4}
\usepackage{float}
\usepackage{mathrsfs}
\usepackage{amsfonts}
\usepackage{array}
\usepackage{epsfig}
\usepackage{amsmath}    
\usepackage{amssymb}   
\usepackage{graphicx}   
\usepackage{verbatim}   
\usepackage{color}      
\usepackage{subfigure}  

\begin{document}

\title{
CT14QED PDFs from Isolated Photon
Production in Deep Inelastic Scattering}

\author{Carl Schmidt}
\email{schmidt@pa.msu.edu}
\author{Jon Pumplin}
\email{pumplin@pa.msu.edu}
\author{ Daniel Stump}
\email{stump@pa.msu.edu}
\author{ C.--P. Yuan}
\email{yuan@pa.msu.edu}
\affiliation{
Department of Physics and Astronomy, Michigan State University,\\
 East Lansing, MI 48824 U.S.A. }

\begin{abstract}
We describe the implementation of Quantum Electrodynamic (QED)
evolution at Leading Order (LO) along with Quantum Chromodynamic (QCD)
evolution at Next-to-Leading Order (NLO) in the CTEQ-TEA Global analysis
package.  The inelastic contribution to the photon Parton Distribution Function (PDF) is described by
a two-parameter ansatz, coming from radiation off the valence quarks, and based on the CT14 NLO PDFs. 
Setting the two parameters to be equal allows us to completely specify the inelastic photon PDF in
terms of the inelastic momentum fraction carried by the photon, $p_0^\gamma$, at the initial scale $Q_0=1.295$ GeV.
We obtain constraints on the photon PDF by comparing with ZEUS data~\cite{Chekanov:2009dq}  on
the production of isolated photons in deep inelastic scattering, $ep\rightarrow e\gamma+X$.
For this comparison we present a new perturbative calculation of the process that consistently combines
the photon-initiated contribution with the quark-initiated contribution. Comparison with the
data allows us to put a constraint at the 90\% confidence level of $p_0^\gamma\lesssim0.14\%$ for the inelastic photon PDF 
at the initial scale of $Q_0=1.295$ GeV in the one-parameter radiative ansatz.  The resulting inelastic CT14QED PDFs will be 
made available to the public. In addition, we also provide CT14QEDinc PDFs, in which the {\em inclusive} photon PDF at the scale $Q_0$ is
defined by the sum of the inelastic photon PDF and the elastic photon distribution obtained from the Equivalent Photon Approximation.


\end{abstract}

\pacs{ 12.38.Bx, 12.38.Cy, 13.40.Ks, 13.85.Qk, 14.70.Bh}

\keywords{parton distribution functions; QED; photon; DIS}

\maketitle

\section{Introduction}\label{sec:intro}

The high precision of current collider data requires comparable precision in the phenomenological predictions.  
The state of the art in high-energy calculations is at next-to-next-to-leading order (NNLO) in Quantum ChromoDynamics (QCD).  Consequently, major efforts have been undertaken to produce NNLO parton distribution functions (PDFs) from a global analysis
of the available data.  These include the CT14NNLO PDFs~\cite{Dulat:2015mca} as well as others~\cite{Ball:2014uwa,Harland-Lang:2014zoa,Alekhin:2013nda}, all of which include LHC data in the determination of the PDFs.

In this paper we describe the introduction of QED evolution at leading order (LO) with the next-to-leading order (NLO) QCD evolution in
the same CTEQ global analysis package that was used to produce the CT14 PDFs~\cite{Dulat:2015mca}.  
Past studies of QED effects in global analysis have been done by the MRST~\cite{Martin:2004dh} and the NNPDF~\cite{Ball:2013hta} groups.  We have checked our code against other QED$+$QCD evolution codes~\cite{Roth:2004ti,Bertone:2013vaa} and find good agreement.

The MRST and NNPDF analyses used different approaches for modeling the photon PDF.  The MRST group used a parametrization for the photon PDF based on radiation off of  ``primordial'' up and down quarks, with the photon radiation cut off at low scales by constituent or current quark masses~\cite{Martin:2004dh}.  The NNPDF group used a more general photon parametrization, which was then constrained by high-energy $W$, $Z$ and Drell-Yan data at the LHC~\cite{Ball:2013hta}.  They found constraints on the size of the photon PDF, which was still consistent with zero at the initial scale of $\sqrt{2}$ GeV.
As discussed by Martin and Ryskin~\cite{Martin:2014nqa}, the photon PDF has a large {\em elastic} contribution in which the proton remains intact, in addition to
the {\em inelastic} contribution in which the proton breaks into a multihadron final state.\footnote{In Ref.~\cite{Martin:2014nqa} these two contributions are referred to as ``coherent'' and ``incoherent'', respectively.}  Neither MRST nor NNPDF addresses these separate contributions to the photon PDF, although we can assume that
the NNPDF photon is {\em inclusive}, containing both inelastic and elastic components, since it was constrained using inclusive Drell-Yan and vector boson data.

Given the limited amount of data to constrain the shape of the photon PDF, we will use a generalization of the MRST approach. 
We parametrize the inelastic contribution to the photon at the initial scale\footnote{The initial scale $Q_0=1.295$ GeV is the same as that used in the standard CT14 PDF sets, and was
chosen to be just below the input charm pole mass of $m_c=1.3$ GeV.} $Q_0=1.295$ GeV by
\begin{equation}
f_{\gamma/p}(x,Q_0)\ =\ \frac{\alpha}{2\pi}\left(A_ue_u^2\tilde{P}_{\gamma q}\circ u^0(x)+A_de_d^2\tilde{P}_{\gamma q}\circ d^0(x)\right)\ ,\label{eq:photonPDFintro}
\end{equation}
where $\tilde{P}_{\gamma q}\circ f^0(x)$ is the convolution of the quark-to-photon splitting function $\tilde{P}_{\gamma q}(x)$
with the ``primordial''
quark distribution $f^0(x)$, which we take to be the initial CT14 NLO up and down valence distributions.
We then set $A_u=A_d$ to obtain a single parameter family of photon distributions, which we can label by their initial inelastic momentum
fraction $p_\gamma^0$.  For comparison, in analogy with the MRST approach, we will also show results for a ``Current Mass'' (CM)
photon distribution, given by defining $A_i=\ln\left(Q_0^2/Q_i^2\right)$, and setting the $Q_i$ to the quark current masses; {\it i.e.,} $Q_u=m_u=6$ MeV and $Q_d=m_d=10$ MeV.

We will constrain the inelastic photon PDF using data on Deep Inelastic Scattering (DIS) with isolated photons from the
ZEUS collaboration~\cite{Chekanov:2009dq}.  The advantage of using this process is that the initial-state photon contributions are at leading
order in the perturbation expansion.  In contrast, the initial-state photon contribution to Drell-Yan or $W$ and $Z$ production
is suppressed by factors of $(\alpha/\alpha_s)$ relative to the leading quark-antiquark production.  However, to use the DIS-plus-photon
data we will first need to address some technical issues relating to the combination of different subprocess contributions to the observed final state.

The organization of this paper is as follows:  In Sec.~\ref{sec:QEDEvol} we describe the inclusion of QED evolution in the CTEQ 
global analysis code and give more details about our initial PDF parametrizations.  In Sec.~\ref{sec:constraints} we discuss
constraints on the photon PDF coming from the CT14 global analysis data set and from the ZEUS DIS with isolated photon data.
In this section we present a new calculation for the DIS-plus-isolated-photon process, which consistently combines the photon-initiated 
contribution with the quark-initiated contribution.  We show that this data gives significant constraints on the initial photon PDF.
In Sec.~\ref{sec:conclude} we discuss our findings and give conclusions.  We also include an Appendix where we show comparisons
between our QCD$+$QED evolution code and other publicly-available codes.

\section{Incorporation of QED effects in CTEQ-TEA global analysis}\label{sec:QEDEvol}

In this section we discuss the implementation of the QED evolution and the initial photon PDF
in the context of the CTEQ-TEA global analysis program.

\subsection{QCD-plus-QED evolution}\label{sec:QCDplusQED}

The evolution of the PDFs, $f(x,\mu_F)$, including QED contributions at leading order (LO) and QCD contributions at
higher orders, is described by the equations:
\begin{eqnarray}
\frac{df_{q_i}}{dt}&=&\frac{\alpha_s}{2\pi}\Biggl(\sum_j\left(P_{q_iq_j}\circ f_{q_j}+P_{q_i\bar{q}_j}\circ f_{\bar{q}_j}\right)+P_{qg}\circ f_g\Biggr)\nonumber\\
&&+
\frac{\alpha}{2\pi} e_{i}^2\Biggl(\tilde{P}^{(0)}_{qq}\circ f_{q_i}+\tilde{P}^{(0)}_{q\gamma}\circ f_\gamma\Biggr)
\nonumber\\
\frac{df_{\bar{q}_i}}{dt}&=&\frac{\alpha_s}{2\pi}\Biggl(\sum_j\left(P_{\bar{q}_i\bar{q}_j}\circ f_{\bar{q}_j}+P_{\bar{q}_iq_j}\circ f_{q_j}\right)+P_{qg}\circ f_g\Biggr)\nonumber\\
&&+
\frac{\alpha}{2\pi} e_{i}^2\Biggl(\tilde{P}^{(0)}_{qq}\circ f_{\bar{q}_i}+\tilde{P}^{(0)}_{q\gamma}\circ f_\gamma\Biggr)
\nonumber\\
\frac{df_{g}}{dt}&=&\frac{\alpha_s}{2\pi}\Biggl(P_{gg}\circ f_{g}+\sum_i P_{gq}\circ  (f_{q_i}+f_{\bar{q}_i})\Biggr)\\
\frac{df_{\gamma}}{dt}&=&\frac{\alpha}{2\pi}\left(\tilde{P}^{(0)}_{\gamma\gamma}\circ f_{\gamma}+\sum_i e_{i}^2\tilde{P}^{(0)}_{\gamma q}\circ 
(f_{q_i}+f_{\bar{q}_i})\right)\ ,\nonumber
\end{eqnarray}
where $t=\ln\mu_F^2$, the indices $i$ and $j$ run over active quark flavors, and the convolution is defined by
\begin{eqnarray}
\bigl(P_{ab}\circ f_b\bigr)(x,\mu_F)&=&\int_0^1dz\int_0^1dy\, \delta(zy-x)\,P_{ab}(z)\, f_b(y,\mu_F)
\ .
\end{eqnarray}
The QCD splitting functions, given by
\begin{equation}
P_{ab}\ =\ \sum_n\left(\frac{\alpha_s}{2\pi}\right)^nP_{ab}^{(n)}\ ,
\end{equation}
are known up to $n=2$, next-to-next-to-leading order (NNLO)~\cite{Moch:2004pa,Vogt:2004mw}.
The LO QED splitting functions can be extracted from the LO QCD splitting functions, giving
\begin{eqnarray}
\tilde{P}^{(0)}_{qq}(z)&=&\frac{1+z^2}{(1-z)_+}+\frac{3}{2}\delta(1-z)\nonumber\\
\tilde{P}^{(0)}_{q\gamma }(z)&=&N_c\left[z^2+(1-z)^2\right]\nonumber\\
\tilde{P}^{(0)}_{\gamma q}(z)&=&\frac{1+(1-z)^2}{z}\\
\tilde{P}^{(0)}_{\gamma \gamma}(z)&=&-\frac{2}{3}N_c\sum_i e_i^2\,\delta(1-z)\ ,
\end{eqnarray}
where $N_c=3$ is the number of colors.

We have modified the Fortran NLO evolution code {\tt evolve}, which was used for previous CTEQ-TEA
 PDFs (CTEQ6-6.6~\cite{Pumplin:2002vw}-\cite{Nadolsky:2008zw} and CT09~\cite{Pumplin:2009nk}), to include
the LO QED contributions.  This code solves the evolution equations directly in $x$-space, so that the only new technical issue introduced by the QED corrections is the separation of the quark singlet distributions into separate up and down contributions, based on the quark charges. We have checked our evolution code against the public QCD$+$QED codes, 
{\tt partonevolution}~\cite{Weinzierl:2002mv,Roth:2004ti} and {\tt APFEL}~\cite{Bertone:2013vaa}, and we find good agreement.
Details of this comparison are given in the Appendix.

\subsection{Initial Photon PDFs}\label{sec:photonPDF}

The initial photon PDF at the scale $Q_0$ is a nonperturbative input that must be obtained by a fit to data.  
Even a choice of zero initial photon PDF is ambiguous, since it depends on the arbitrary scale $Q_0$.  
So far there have been two different approaches to the initial photon PDF.  In the 
MRST analysis~\cite{Martin:2004dh} the initial photon PDF was given by an ansatz, obtained from radiation off primordial
valence up and down quark distributions, cut off at low scales given by the current quark masses, $m_u=6$ MeV and $m_d=10$ MeV, or by
constituent quark masses $m_U=m_D=300$ MeV.  Alternatively, the NNPDF approach~\cite{Ball:2013hta} was to use a general parametrization
for the initial photon PDF to be constrained by high-energy $W$, $Z$, and Drell-Yan production at the LHC.

In this work we will use a generalization of the MRST ansatz, but we must first address a subtlety of the photon PDF.  Unlike the case for colored partons, the photon PDF has a large elastic component, in which the proton remains intact~\cite{Martin:2014nqa}.  This is in addition to the inelastic component, in which
the proton dissociates.  The elastic component can be parametrized by the Equivalent Photon Approximation (EPA)~\cite{Budnev:1974de}, which involves an integration over the proton electromagnetic form factors.  For this work we focus on the inelastic component, which we parametrize by a radiative ansatz, but with free parameters to be fit by data.  Given the weak constraints from data on the photon PDF, we find it useful to limit the number of parameters to one or two
for the time being.  We shall see that the ZEUS DIS-plus-isolated-photon data~\cite{Chekanov:2009dq} constrains the inelastic photon PDF roughly in the range $10^{-3}< x<2\cdot10^{-2}$ for $16< Q^2 <300$ GeV$^2$.

We parametrize the inelastic contribution to  the initial photon PDFs in the proton and neutron by
\begin{eqnarray}
f_{\gamma/p}(x,Q_0)&=&\frac{\alpha}{2\pi}\left(A_ue_u^2\tilde{P}_{\gamma q}\circ u^0(x)+A_de_d^2\tilde{P}_{\gamma q}\circ d^0(x)\right)\nonumber\\
f_{\gamma/n}(x,Q_0)&=&\frac{\alpha}{2\pi}\left(A_ue_u^2\tilde{P}_{\gamma q}\circ d^0(x)+A_de_d^2\tilde{P}_{\gamma q}\circ u^0(x)\right)
\ ,\label{eq:photonPDF}
\end{eqnarray}
where $u^0$ and $d^0$ are ``primordial'' valence-type distributions in the proton, and the initial photon PDF in the neutron is
obtained by an approximate isospin symmetry.  Defining  $A_i=\ln\left(Q_0^2/Q_i^2\right)$, we can trade the parameters $A_i$ for mass scales $Q_i$, and we see that the nonperturbative inputs $f_{\gamma/(p,n)}(x,Q_0)$ are modeled by the radiation of a single photon off the ``primordial'' quarks, with a collinear cutoff given by the scales $Q_i$.  The MRST initial photon PDFs can be obtained from this parametrization by setting $Q_0=1$ GeV, using the
functions for $u_0$ and $d_0$ given in Ref.~\cite{Martin:2004dh}, and letting $Q_u$ and $Q_d$ be either the constituent or current
quark masses.  For our analysis, we use $Q_0=1.295$ GeV as in CT14, and we set 
\begin{eqnarray}
u_0(x)&=&u^p_V(x,Q_0)\ =\ f_{u/p}(x,Q_0^2)-f_{\bar{u}/p}(x,Q_0^2)\nonumber\\
d_0(x)&=&d^p_V(x,Q_0)\ =\ f_{d/p}(x,Q_0^2)-f_{\bar{d}/p}(x,Q_0^2)
\ ,
\end{eqnarray}
the initial up and down CT14 NLO valence distributions in the proton.  

The presence of the photon PDF violates isospin between the neutron and proton.  Continuing with the radiative ansatz
and working to first order in $\alpha$, we can neglect the isospin violation in the gluon and sea-quark PDFs ~\cite{Martin:2004dh} and use
\begin{eqnarray}
f_{g/p}(x,Q_0)&=&f_{g/n}(x,Q_0)\nonumber\\
f_{\bar{q}/p}(x,Q_0)&=&f_{\bar{q}/n}(x,Q_0) \qquad\mbox{for}\quad \bar{q}=\bar{u},\bar{d},\bar{s},\bar{c},\bar{b}\label{eq:sea}\\
f_{{q}/p}(x,Q_0)&=&f_{{q}/n}(x,Q_0)\qquad \mbox{for}\quad q=s,c,b\ .\nonumber
\end{eqnarray}
For the valence quarks at first order in
$\alpha$, the radiative ansatz plus approximate isospin symmetry implies
\begin{eqnarray}
u^p_V(x,Q_0)&\approx&u^0(x)+\frac{\alpha}{2\pi}A_ue_u^2\tilde{P}_{q q}\circ u^0(x)\nonumber\\
d^p_V(x,Q_0)&\approx&d^0(x)+\frac{\alpha}{2\pi}A_de_d^2\tilde{P}_{q q}\circ d^0(x)\nonumber\\
u^n_V(x,Q_0)&\approx&d^0(x)+\frac{\alpha}{2\pi}A_ue_u^2\tilde{P}_{q q}\circ d^0(x)\label{eq:valenceISO}\\
d^n_V(x,Q_0)&\approx&u^0(x)+\frac{\alpha}{2\pi}A_de_d^2\tilde{P}_{q q}\circ u^0(x)\nonumber
\ .
\end{eqnarray}
This suggests a consistent set of PDFs for the valence quarks in the neutron given by
\begin{eqnarray}
u^n_V(x,Q_0)&=&d^p_V(x,Q_0)+\frac{\alpha}{2\pi}\left(A_ue_u^2-A_de_d^2\right)\tilde{P}_{q q}\circ d^0(x)\nonumber\\
d^n_V(x,Q_0)&=&u^p_V(x,Q_0)+\frac{\alpha}{2\pi}\left(A_de_d^2-A_ue_u^2\right)\tilde{P}_{q q}\circ u^0(x)
\ .\label{eq:valenceN}
\end{eqnarray}
Note that Eqs.~(\ref{eq:photonPDF}) and (\ref{eq:valenceN})  together ensure that if the number and momentum sum
rules (including the photon contribution) are satisfied for the PDFs in the proton, they are automatically satisfied for the
PDFs in the neutron\footnote{This simple approximate isospin symmetry is broken by the inclusion of the elastic component of
the photon PDF, since there is no corresponding elastic photon in the neutron.}, regardless of the choices for $u^0$ and $d^0$.  
Again, for our analysis, we choose $u_0$ and $d_0$ to equal the initial up and down CT14 NLO valence distributions in the proton.
Thus, from Eqs.~(\ref{eq:photonPDF})-(\ref{eq:sea}) and (\ref{eq:valenceN}) we can obtain the quark, gluon, and photon PDFs in both the proton and neutron in terms of the parametrization
of the quark and gluon PDFs in the proton, plus the two additional parameters $A_u$ and $A_d$.

\begin{figure}[t]
\begin{center}
\includegraphics[width=0.47\textwidth]{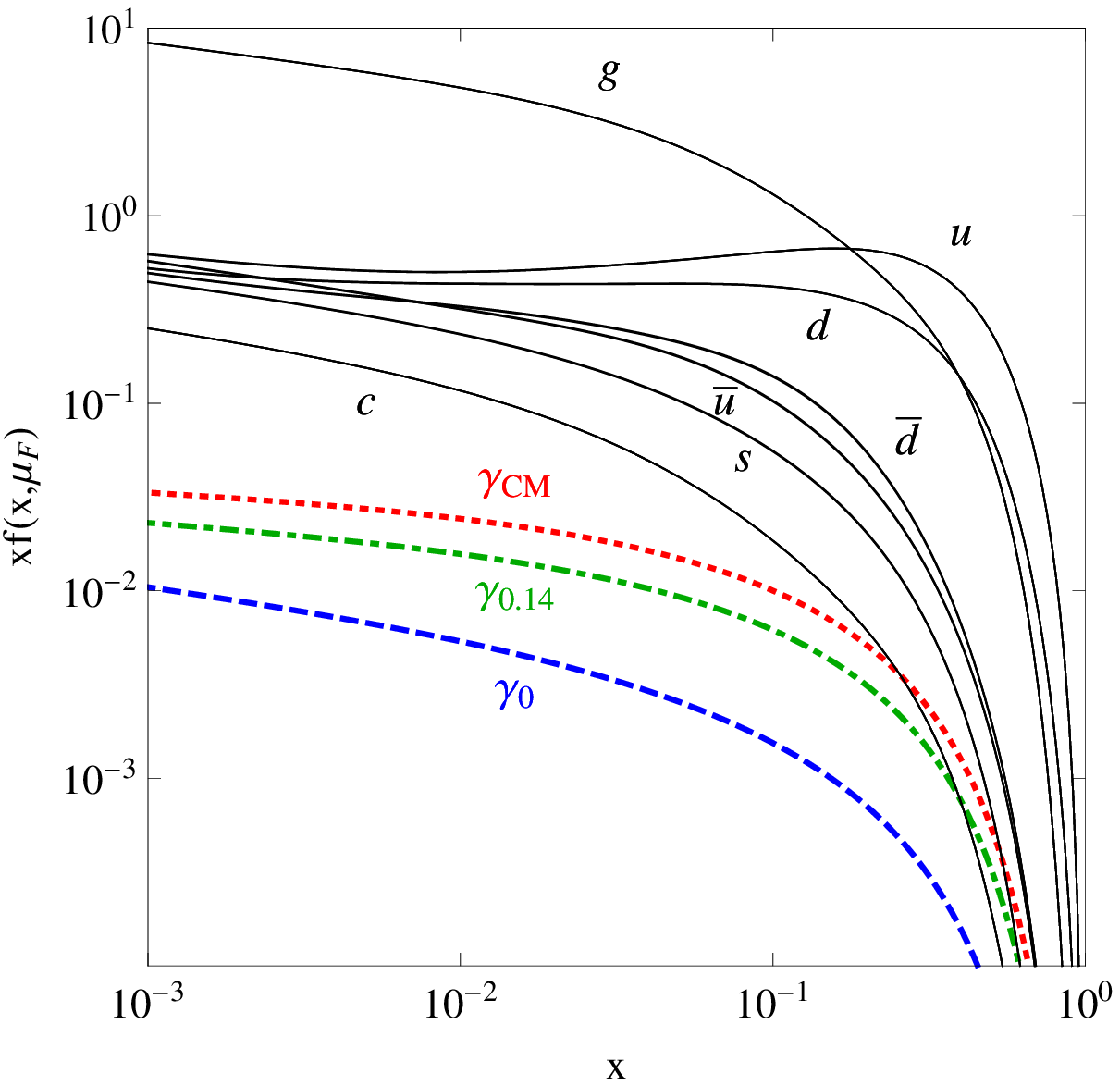}
\includegraphics[width=0.47\textwidth]{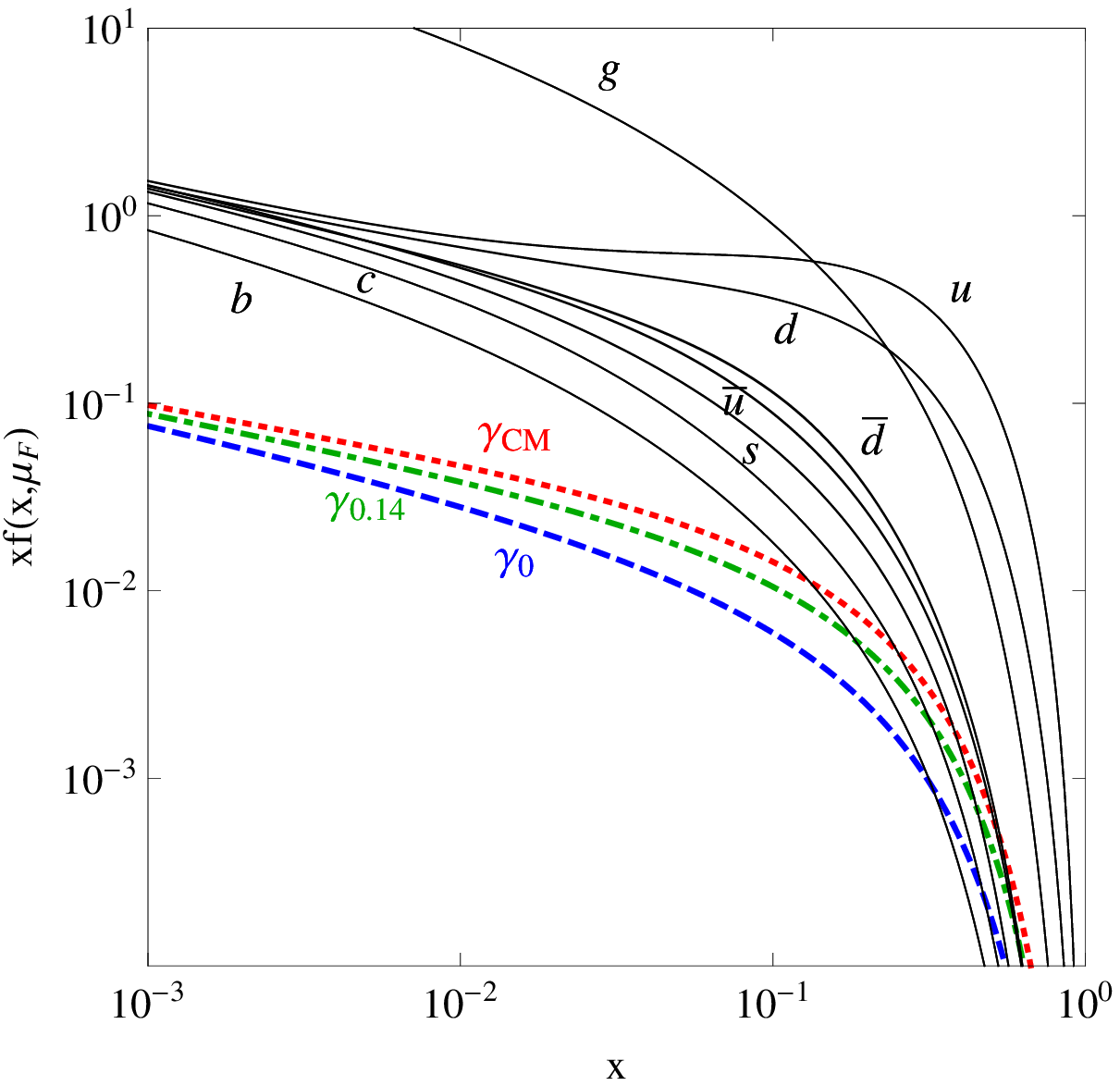}
\end{center}
\caption{Plots of $xf(x,\mu_F)$ for $\mu_F=3.2$ GeV (left) and $\mu_F=85$ GeV (right).
Three representative photon PDFs are plotted: the ``Current Mass'' photon PDF ($\gamma_{\rm CM}$, red dotted), and photon PDFs with initial inelastic photon momenta fractions of $p_0^\gamma=0$ and 0.14\% ($\gamma_{0}$, blue dashed, and $\gamma_{0.14}$, green dot-dashed, respectively).  The effects of the different initial photon PDFs on the quark and gluon PDFs are imperceptible in these plots.
\label{fig:xfQ}}
\end{figure}

For this paper, we shall make the further simplification of $A_u=A_d$, which corresponds to cutting off the radiation
from both valence quarks at the same scale.  With this choice, everything is specified
by one additional parameter, which can be taken to be the cutoff scale $Q_{\rm cut}=Q_u=Q_d$, defined by $A_u=A_d
=\ln\left(Q_0^2/Q_{\rm cut}^2\right)$.  Alternatively, we can trade this parameter for the initial inelastic photon momentum fraction 
in the proton:
\begin{eqnarray}
p_0^\gamma&=&\int_0^1 dx\,xf_{\gamma/p}(x,Q_0)\ .
\end{eqnarray}
For the remainder of this paper, unless otherwise specified, the photon PDFs will be in this one-parameter radiation ansatz
labeled by $p_0^\gamma$.
For comparison purposes, we will make one exception to this by defining a  ``Current Mass'' (CM) photon PDF, 
analogous to the MRST current mass PDF, and given by
$A_i=\ln\left(Q_0^2/Q_i^2\right)$ with $Q_u=m_u=6$ MeV and $Q_d=m_d=10$ MeV.  For this choice the initial inelastic photon momentum
fraction is determined to be $p_0^\gamma=0.26$\%.
For all other partons in our analysis we use the CT14 NLO initial distributions, except that to maintain a total momentum fraction of 1, 
we re-normalize the initial up, down, and strange sea-quark distributions, to account for the additional photon momentum fraction.
Given that the relevant photon momentum fractions are very small, we find that this reduces the sea-quark distributions by typically less
than 1\%, and it is inconsequential in our analysis.  (The sea-quark distributions are reduced by 0.9\% for $p_0^\gamma=0.14$\%, and they are reduced by 1.6\% for the CM photon PDF.)
In Fig.~\ref{fig:xfQ} we plot the quantity $xf(x,\mu_F)$ for three representative
photon PDFs, relative to the quark and gluon PDFs, at the scales $\mu_F=3.2$ GeV and $\mu_F=85$ GeV.
We note that the effect of the initial photon PDF and the QED evolution on the quark and gluon PDFs is imperceptible in these 
plots.  For the photon PDF, we plot for $p_0^\gamma=0$\% ($Q_{\rm cut}=Q_0=1.295$ GeV) and for
$p_0^\gamma=0.14$\% ($Q_{\rm cut}=71$ MeV), and for the CM photon PDF.

\section{Constraints on the Photon PDF}\label{sec:constraints}

\subsection{Constraints from the CT14 data set}\label{sec:ct10}

\begin{figure}[t]
\begin{center}
\includegraphics[width=0.47\textwidth]{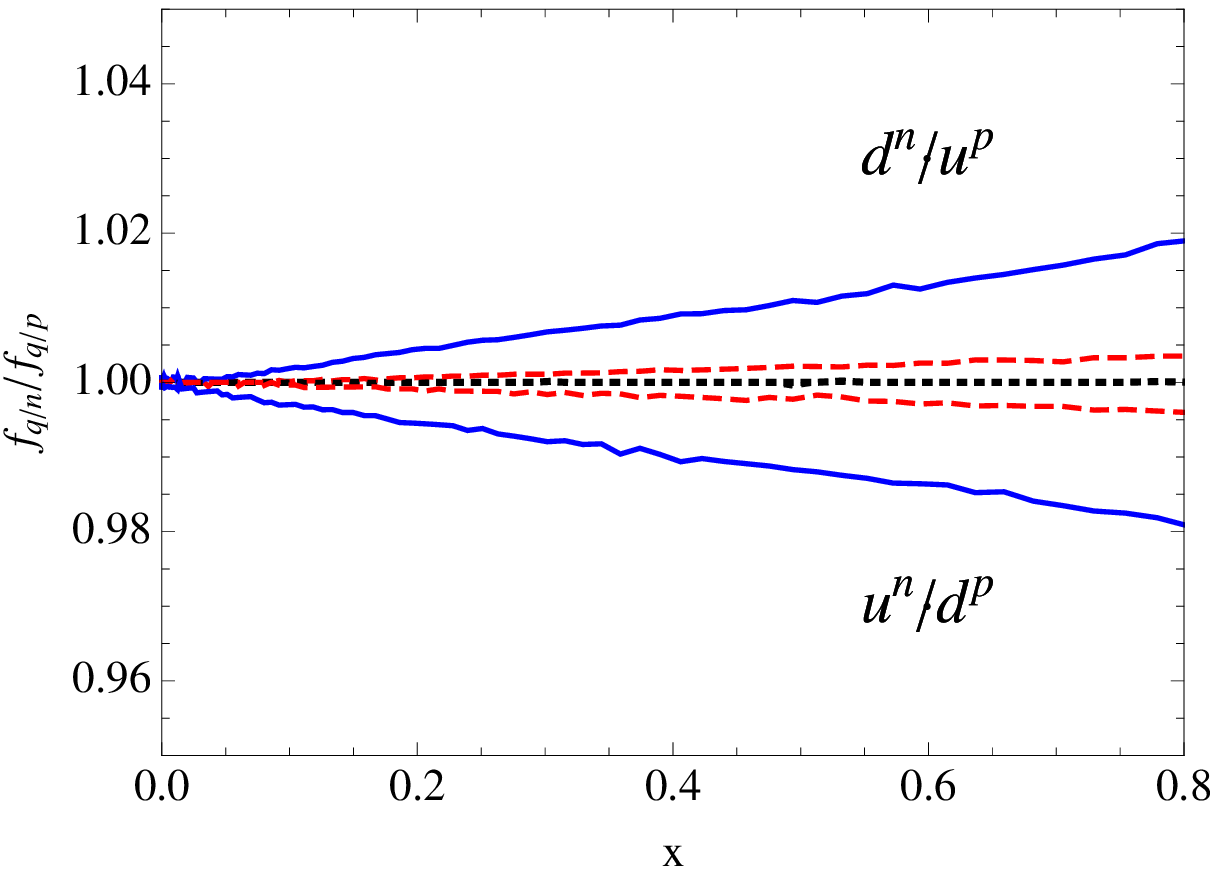}
\includegraphics[width=0.47\textwidth]{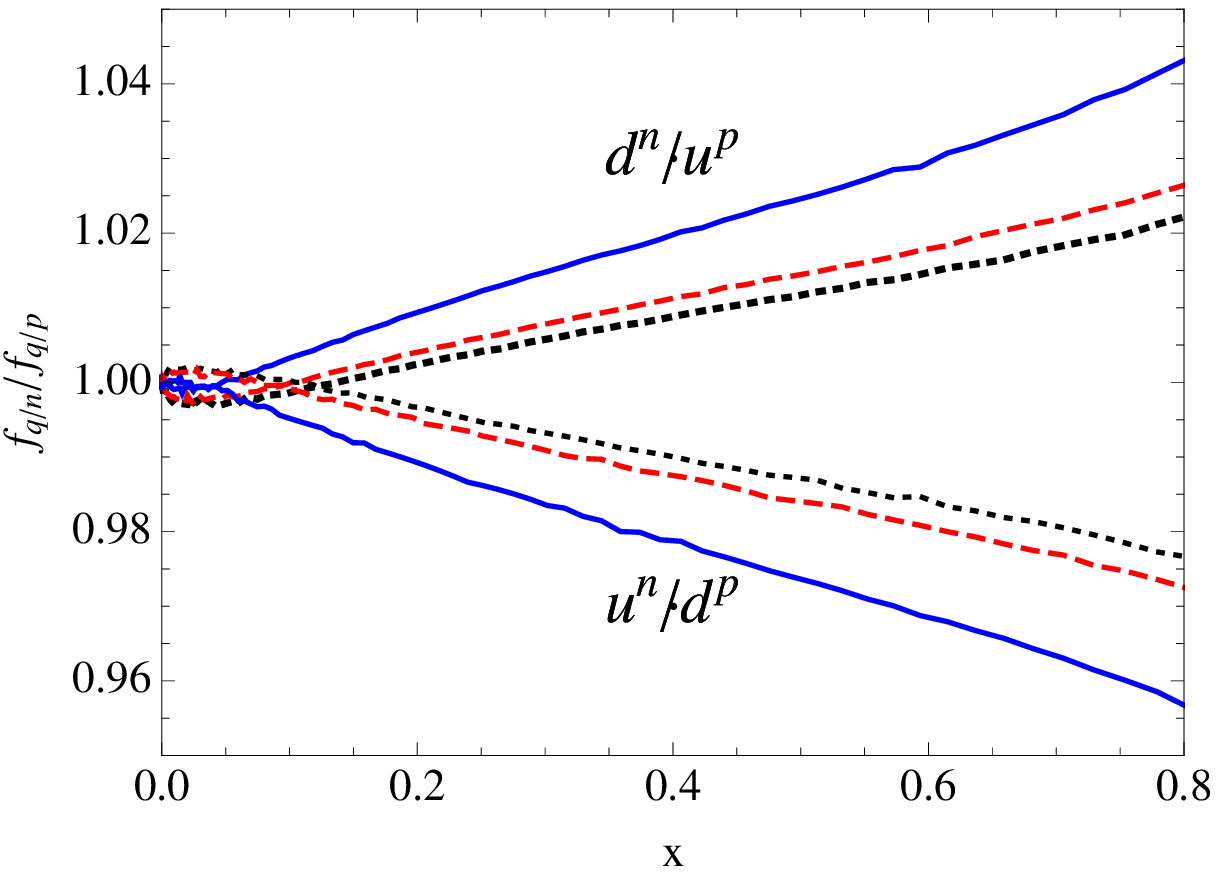}
\end{center}
\caption{Plots of $f_{d/n}(x,\mu_F)/f_{u/p}(x,\mu_F)$  and $f_{u/n}(x,\mu_F)/f_{d/p}(x,\mu_F)$ 
for $\mu_F=1.3$ GeV (black dots), 3.2 GeV (red dashes), and 85 GeV (blue solid).
The left plot is for zero initial photon momentum and the right plot is for the CM photon PDF.
\label{fig:isospin}}
\end{figure}

The constraints on the photon PDF from the DIS and Tevatron  data, used in the CT14 analysis, are relatively
weak.  These come from two main sources: isospin violation effects in nuclear scattering 
and constraints from the momentum sum rule.
In general, isospin violation will arise through QED evolution,
as well as from the initial conditions given by Eq.~(\ref{eq:valenceN}).  This isospin violation can be seen in Fig.~\ref{fig:isospin}, where we plot
$f_{d/n}(x,\mu_F)/f_{u/p}(x,\mu_F)$  and $f_{u/n}(x,\mu_F)/f_{d/p}(x,\mu_F)$ for several values of $\mu_F$ for
the case where the initial photon PDF is zero, and for the case where the initial photon is the CM choice.  
Note that the isospin violation is small and most important at large $x$.  Given that cuts of $W^2=Q^2(1/x-1)>12$ GeV$^2$,
applied to enforce perturbativity in the calculations, typically require $x\lesssim0.2-0.4$, we expect constraints from
isospin violation to be small in the present data, as observed in the MRST analysis~\cite{Martin:2004dh}.  

Constraints from the momentum
sum rule arise because any momentum carried by the photon implies less momentum available for the
quark and gluon PDFs.  In this way, constraints from data on the colored parton PDFs indirectly impact the
photon PDF.  We have performed a preliminary analysis using the data sets included for CT10~\cite{ct10nn}.  For a fixed initial photon momentum fraction,
with the photon PDF parametrized as discussed in Sec.~II, we minimized the global $\chi^2$ by varying the quark and gluon PDFs.
Using the usual CTEQ-TEA choice of $\Delta\chi^2<100$ tolerance, we obtain a limit on the photon momentum fraction of $p_0^\gamma<5.6$\% at the 90\%
confidence level, which is similar in magnitude to the results found by the MRST and NNPDF analyses.  The best fit for the initial photon momentum 
fraction from this global analysis is $p_0^\gamma=1.2$\%, but with only a small change of $\Delta\chi^2=-7$, relative to the fit with $p_0^\gamma=0$\%.
For comparison, we find the elastic contribution to the initial photon momentum fraction, as calculated in the equivalent photon approximation, to be
$p_{0,\,{\rm elastic}}^\gamma=0.15$\%.

Unfortunately, this limit on the initial photon momentum fraction is much larger than one would expect for a photon PDF.
In the analysis of the NNPDF group, additional constraints were made on the initial photon PDF by including LHC data on
high-energy $W$, $Z$, and Drell-Yan production, and comparing with theoretical predictions that included photon initial-state contributions.
 Although the photon-induced contribution to these processes is small compared to the dominant quark-antiquark annihilation subprocess, 
 the precision of these measurements was enough to substantially increase the constraints on the photon PDF~\cite{Ball:2013hta}.  
However, the small relative contribution of the photon-photon subprocess puts a stringent requirement on the precision needed for both experimental and
theoretical analyses.  Any small misjudgment of systematic errors on the experimental side, or uncalculated higher-order corrections on the theoretical side
could have a significant effect on the extraction of the photon PDF.  In particular, given that the initial photon PDF is nominally of order $\alpha$, one might
expect that the uncalculated ${\cal O}(\alpha^2)$ quark-initiated contributions to Drell-Yan production would contribute at the same level as the photon-initiated
contributions.  For this reason, we consider a different experimental process, isolated photon production in DIS,
to constrain the photon PDF.  

\subsection{Calculation of the process $ep\rightarrow e\gamma+X$}\label{sec:theory}

\begin{figure}[t]
\begin{center}
\includegraphics[width=0.8\textwidth]{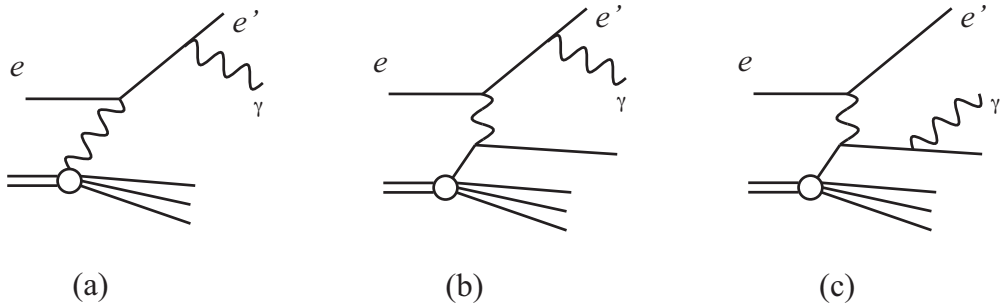}
\end{center}
\caption{Amplitudes for the process $ep\rightarrow e\gamma+X$.  For each diagram shown there is
an additional diagram where the photon is emitted off the initial-state lepton or quark.
 \label{fig:Feyndiags}}
\end{figure}

At the partonic level, the process of DIS with isolated photon occurs at LO through Compton scattering
of a photon coming from the proton off the lepton, as shown in Fig.~\ref{fig:Feyndiags}(a).  
Thus, this process probes the photon PDF at LO, having no large backgrounds with which to compete.  However, the quark-initiated subprocess shown in Figs.~\ref{fig:Feyndiags}(b) and (c), while formally suppressed by
${\cal O}(\alpha)$, is just as large because of the small size of $f_\gamma$ relative to $f_q$.  In fact, if we consider
the photon PDF to be ${\cal O}(\alpha)$, then the photon-initiated subprocess and the quark-initiated subprocess are actually
the same order in $\alpha$.  Thus, the correct way to calculate the cross section for DIS with isolated photons is to include
both subprocesses consistently without double-counting.

In the literature there have been two approaches to calculations of the process $ep\rightarrow e\gamma+X$.  The calculation of 
MRST~\cite{Martin:2004dh}, which was preceded by studies of Bl\"umlein {\it et al.}~\cite{Blumlein:1989gk,Blumlein:1993ef,Arbuzov:1995id}, included just the 
photon-initiated contribution of Fig.~\ref{fig:Feyndiags}(a).  The calculation
of Gehrmann-De Ridder, Gehrmann, and Poulsen (GGP)~\cite{GehrmannDe Ridder:2006wz,GehrmannDeRidder:2006vn}
included just the quark-initiated contributions of Figs.~\ref{fig:Feyndiags}(b,c).  
In the GGP analysis it was found convenient
to make the Lorentz-invariant separation of the cross section into three components, depending on the fermion line off which
the final-state photon is emitted:  $LL$ for emission off the lepton line, given by the square of the partonic amplitude in Fig.~\ref{fig:Feyndiags}(b); $QQ$ for emission off the quark line, given by the square of the partonic amplitude in Fig.~\ref{fig:Feyndiags}(c); and $QL$ for the interference between the two sets of diagrams.\footnote{Note that each of the diagrams in Fig.~\ref{fig:Feyndiags} actually represents two Feynman diagrams, where the
final-state photon is emitted off the initial-state lepton or quark as well as off the final-state lepton or quark.}
In the GGP calculation a cut on the outgoing quark was necessary to remove the divergence in the amplitude as the photon off-shellness went
to zero in the $LL$ amplitude.   A hybrid calculation was also considered by the ZEUS Collaboration in their analysis of the DIS-plus-isolated photon data~\cite{Chekanov:2009dq}, where the $LL$ component
of the quark-initiated subprocess of GGP was replaced by the photon-initiated subprocess of MRST.

In this section we introduce a consistent and systematic method of combining the photon- and quark-initiated subprocesses,
which also reduces the factorization scale dependence of either calculation.  First, consider the calculation of the differential 
cross section as a power series
in $\alpha$ without consideration of the relative sizes of $f_\gamma$ and $f_q$.  It can be written
as a convolution over partonic cross sections
\begin{equation}
d\sigma\,=\,\sum_a \int_0^1 d\xi \,f_a(\xi,\mu_F) d\hat\sigma_a\ ,
\end{equation}
where each of the partonic cross sections can be expanded in a power series in $\alpha$,
\begin{equation}
d\hat\sigma_{a}\ =\ \sum_n d\hat\sigma_{a}^{(n)}\ .
\end{equation}
The diagram in Fig.~\ref{fig:Feyndiags}(a)
corresponds to a LO contribution ($a=\gamma;\,n=0$), while the diagrams in Figs.~\ref{fig:Feyndiags}(b,c) correspond to NLO contributions ($a=q,\bar{q};\,n=1$).
Through NLO in $\alpha$ the cross section can still be written as a sum of $LL$, $QQ$, and $QL$ components,
\begin{equation}
d\sigma\,=\,d\sigma^{(LL)}+d\sigma^{(QQ)}+d\sigma^{(QL)}\ ,
\end{equation}
where the $LL$ component also includes the photon-initiated contribution,
\begin{equation}
d\sigma^{(LL)}\,=\,\int_0^1 d\xi \left[f_\gamma(\xi,\mu_F) d\hat\sigma_\gamma+\sum_{a=q,\bar{q}} f_a(\xi,\mu_F) d\hat\sigma^{(LL)}_a\right]\ ,
\end{equation}
Using the modified minimal subtraction ($\overline {\rm MS}$) scheme,
we can factorize the initial-state singularity from the NLO quark-initiated subprocess into the definition of the photon PDF, 
leaving a NLO subprocess cross section,
\begin{equation}
d\hat\sigma_{q}^{(1,LL)}\ =\ d\sigma_{q}^{(1,LL)}+
\frac{\alpha}{2\pi}\left(\frac{4\pi\mu_R^2}{\mu_F^2}\right)^\epsilon\frac{1}{\epsilon \Gamma(1-\epsilon)}\int_0^1dz\,
\tilde{P}_{\gamma q}(z)\, \,d\hat\sigma_\gamma^{(0)}(z\xi)
\ ,\label{eq:AP}
\end{equation}
where the first term on the right-hand side, $d\sigma_q^{(1,LL)}$, is the hard partonic quark-induced subprocess,
and the second term is the collinear subtraction counterterm.  Here, we have distinguished the renormalization scale
$\mu_R$ from the factorization scale $\mu_F$, and we explicitly note that the initial-state collinear singularity
cancels within the $LL$ component.
Calculating everything in dimensional regularization
with $4-2\epsilon$ dimensions, the combined $LL$ component of the NLO quark-initiated subprocess cross section, $d\hat\sigma_q^{(1,LL)}$ is finite as $\epsilon\rightarrow0$.

In principle, there are additional virtual and real contributions at NLO in $\alpha$, besides the quark-initiated contributions.
However, all other NLO terms are proportional to $f_\gamma$, which is in fact suppressed by an amount of order
$\alpha$ relative to $f_q$, as seen in the previous section.
Thus, by keeping the photon-initiated contribution at LO and the quark-initiated contributions at NLO, and 
including the collinear-subtraction counterterm of Eq.~(\ref{eq:AP}) we have a well-defined calculation in the 
$\overline {\rm MS}$ scheme, while maintaining all contributions of the same size in $\alpha$.  Note that this is reminiscent 
of the ACOT scheme~\cite{Aivazis:1993pi} for including charm and bottom quark PDFs, although for the photon, there is no equivalent of the fixed-flavor scheme, due to its zero mass and  the consequently 
nonperturbative photon PDF.

The calculations of the $QQ$ and $QL$ components are identical to those in the GGP calculation.  For the kinematics of
interest to us, the $QL$ component is negligible, but it is included for completeness.
The $QQ$ contribution has a final-state singularity, when the photon and
final-state quark become collinear, which can be handled by including a fragmentation contribution
and an associated collinear subtraction counterterm in the $\overline {\rm MS}$ scheme.  Thus, we obtain
\begin{eqnarray}
d\hat\sigma_{q}^{(1,QQ)}&=&\ d\sigma_{q}^{(1,QQ)}+
\frac{\alpha}{2\pi}\left(\frac{4\pi\mu_R^2}{\mu_f^2}\right)^\epsilon\frac{1}{\epsilon \Gamma(1-\epsilon)}\int_0^1dz\,
\tilde{P}_{\gamma q}(z)\, \,d\hat\sigma_{eq\rightarrow e^\prime Q^\prime}\Big|_{Q^\prime=k^\prime/z}\nonumber\\
&&+\int_0^1 dz\,D_{\gamma/Q^\prime}(z,\mu_{ f})\,d\hat\sigma_{eq\rightarrow e^\prime Q^\prime}\Big|_{Q^\prime=k^\prime/z}\ ,
\end{eqnarray}
where $d\sigma_{q}^{(1,QQ)}$ is the hard partonic subprocess, $d\hat\sigma_{eq\rightarrow e^\prime Q^\prime}$ is the LO subprocess cross section for $eq\rightarrow e^\prime Q^\prime$, and 
$D_{\gamma/Q^\prime}(z,\mu_F)$ is the fragmentation function for finding the photon (with momentum $k^\prime$) in the quark $Q^\prime$ (with momentum $Q^\prime$), with
momentum fraction $z$ at the fragmentation scale $\mu_f$.  The singularities in the first two terms on the right-hand side of this
equation cancel as $\epsilon\rightarrow0$. We will discuss the choice for the fragmentation function in the next subsection.

We have calculated the $LL$ and the $QQ$ components of the cross section, using the subtraction method to handle
the collinear divergences~\cite{Catani:1996vz} in the hard cross-section term.  Following this method, we subtract a term with a 
two-particle final state
mapped onto a 3-particle final-state, with the third particle phase space unintegrated, and then add back the exact same
term with the third particle phase space integrated out.  The subtraction term is designed to have the same
 collinear-singular limit as the hard term in the same region of phase space of the third particle, so that the hard cross section term minus the subtraction is
integrable in $d=4$ dimensions, while the $1/\epsilon$ singularities in the remaining terms cancel.  Using this method we
obtain for the $LL$ quark-initiated contribution:
\begin{equation}
d\hat\sigma_{q}^{(1,LL)}\ =\ \biggl(d\sigma_{q}^{(1,LL)}-d\sigma_{q\rm (sub)}^{(1,LL)}\biggr)+d\hat\sigma_{q\rm (AP)}^{(1,LL)}\ ,
\label{eq:AP2}
\end{equation}
where
\begin{equation}
d\hat\sigma_{q\rm (AP)}^{(1,LL)}\ =\ 
\int_0^1dz\,\frac{\alpha}{2\pi}
\left[\tilde{P}_{\gamma q}(z)\ln\left(\frac{s\xi(1-z)^2}{\mu_F^2}\right)+e_q^2z\right]\, \,d\hat\sigma_\gamma^{(0)}(z\xi)
\ .\label{eq:AP3}
\end{equation}
The hard term minus the subtraction term can be written as an integral over the phase space of the additional quark:
\begin{eqnarray}
\biggl(d\sigma_{q}^{(1,LL)}-d\sigma_{q\rm (sub)}^{(1,LL)}\biggr)\ =\ \ \ \ \ &&\label{eq:hms}\\
\int_0^1 dw \int_0^{2\pi} \frac{d\phi}{2\pi}&& \!\!\!\!\!\left(2\pi\frac{d^2\left(d\sigma_{q}^{(1,QQ)}\right)}{dw\,d\phi}-
\int_0^1dz\,
\frac{\alpha}{2\pi w}\tilde{P}_{\gamma q}(z)\, \,d\hat\sigma_\gamma^{(0)}(z\xi)\right)
\ ,\nonumber
\end{eqnarray}
where $w=(1-\cos\theta)/2$ with $\theta$ and $\phi$ the scattering angles of the final-state quark in the initial parton-parton
center-of-momentum frame.  Note that the hard term on the right-hand side is treated with three-body final-state phase space, while the
subtraction term is treated with two-body final-state phase space.

Similarly, we obtain for the $QQ$ contribution:
\begin{eqnarray}
d\hat\sigma_{q}^{(1,QQ)}&=&\ \biggl(d\sigma_{q}^{(1,QQ)}-d\sigma_{q\rm (sub)}^{(1,QQ)}\biggr)+d\hat\sigma_{q\rm (frag)}^{(1,QQ)}\ ,
\end{eqnarray}
where
\begin{eqnarray}
d\hat\sigma_{q\rm (frag)}^{(1,QQ)}&=&
\int_0^1 dz\Bigg\{D_{\gamma/Q^\prime}(z,\mu_{ f})\\
&&\qquad\quad\quad+\frac{\alpha}{2\pi}\left[\tilde{P}_{\gamma q}(z)\ln\left(\frac{s\xi z^2(1-z)}{\mu_f^2}\right)+e_q^2z\right]\Bigg\}\,d\hat\sigma_{eq\rightarrow e^\prime Q^\prime}\ .\nonumber
\end{eqnarray}
The hard term minus the subtraction term can be written
\begin{eqnarray}
\biggl(d\sigma_{q}^{(1,QQ)}-d\sigma_{q\rm (sub)}^{(1,QQ)}\biggr)\ =\ \ \ \ \ &&\label{eq:hms2}\\
\int_0^1 d\tilde{w} \int_0^{2\pi} \frac{d\tilde{\phi}}{2\pi}&& \!\!\!\!\!\left(2\pi\frac{d^2\left(d\sigma_{q}^{(1,LL)}\right)}{d\tilde{w}\,d\tilde{\phi}}-
\int_0^1dz\,
\frac{\alpha}{2\pi \tilde{w}}\tilde{P}_{\gamma q}(z)\, \,d\hat\sigma_{eq\rightarrow e^\prime Q^\prime}\right)
\ ,\nonumber
\end{eqnarray}
where in this case we have found it convenient to use a different parametrization of the final-state quark phase space.  Letting $qe\rightarrow q^\prime e^\prime \gamma$ be the hard partonic subprocess, we use $\tilde{w}=(1-\cos\tilde{\theta})/2$, where $\tilde{\theta}$
is the angle between $q^\prime$ and $\gamma$ in the $q^\prime e^\prime$ center-of-momentum frame
and $\tilde{\phi}$ is the azimuthal angle between the $qe \gamma$ plane 
and the $q^\prime e^\prime \gamma$ plane.
As before, the hard term on the right-hand side is treated with three-body final-state phase space, while  in this case the subtraction term
and the fragmentation term are treated in the limit where the final-state photon and quark are collinear, with momenta satisfying $k^\prime=zQ^\prime$ and $q^\prime=(1-z)Q^\prime$.

\subsection{ZEUS Experimental Cuts and Photon Isolation}\label{sec:data}

The ZEUS experiment~\cite{Chekanov:2009dq} used proton and lepton beam energies of $E_p=920$ GeV and $E_\ell=27.5$ GeV, respectively,
corresponding to a center-of-mass energy and rapidity of
\begin{eqnarray}
\sqrt{s}&=&2\sqrt{E_p E_\ell}\ =\ 318\mbox{ GeV}\nonumber\\
Y&=&\frac{1}{2}\ln{\frac{E_p}{E_\ell}}\ =\ 1.76\ ,
\end{eqnarray}
respectively. (We neglect the proton mass, $m_p$, in all calculations here.)
For the process $ep\rightarrow e\gamma+X$,
with momentum satisfying, $\ell+p=\ell^\prime+k^\prime+p^\prime_X$, one can define the
standard DIS variables that describe the kinematics of the scattered lepton, $Q^2=-(\ell-\ell^\prime)^2$, $y=p\cdot(\ell-\ell^\prime)/(p\cdot\ell)$, and $x=Q^2/(sy)$.
The ZEUS Collaboration measured distributions for two leptonic variables $Q$ and $x$, and for two photonic variables,
$E_{\perp\gamma}$ and $\eta_\gamma$, the transverse energy and pseudorapidity of the photon, respectively.
The collaboration combined data that were 59.1\% $e^-p$ and 40.9\% $e^+p$ scattering.  Note that the sign of the charged lepton
has no effect on the $LL$ or $QQ$ components of the calculation, but the combination of the two charged lepton contributions produces a significant cancellation of the already small $QL$ component, so that it is negligible in the analysis.

 The kinematic region defined by the experiment was
\begin{eqnarray}
10 \mbox{ GeV}^2\ <&Q^2&<\ 350 \mbox{  GeV}^2\nonumber\\
&E_\ell^\prime&>\  10\mbox{  GeV}\nonumber\\
139.8^\circ\ <&\theta_\ell^\prime&<\ 171.8^\circ\\ \label{eq:zeuscuts}
4 \mbox{ GeV}\ <&E_{\perp\gamma}&<\ 15 \mbox{  GeV}\nonumber\\
-0.7\ <&\eta_\gamma&<\ 0.9\ .\nonumber
\end{eqnarray}
The cut on the final-state lepton angle, $\theta_\ell^\prime$, can be written
in terms of its rapidity as
\begin{eqnarray}
-2.6355\ <&\eta_\ell^\prime&<\ -1.0053\ .
\end{eqnarray}

There are two additional cuts that require discussion.  The experimentalists reported a cut of
$W_X> 5$  GeV, where $W_X^2=(p+\ell-\ell^\prime-k^\prime)^2$.  Naively, this cut looks problematic 
because it would remove the photon-initiated contribution, which occurs at exactly $W_X=0$.  
However, upon closer investigation it appears that this cut was only applied to the theoretical
and Monte Carlo calculations. To quote from Ref.~\cite{Forrest:2010zza},  `` The keen reader will note that no such cut was applied at the detector level.  This proved impossible due to the poor
 $W_X$ resolution at detector level and poor description of the data by MC...''.  The relevant detector-level cut was
 the requirement of at least one reconstructed track, well separated from the lepton, which was used to
 ensure some hadronic activity and to remove deeply virtual Compton scattering events.  After this cut, it was found that 
 the number of events in the Monte Carlo calculations with $W_X<5$ GeV was negligible.   For our purposes, we interpret the forward track cut as a requirement to tag inelastic events, and we include no explicit $W_X$ cut.
 
 Note that the forward track cut should equally remove elastic isolated photon events, and so remove the contribution from the elastic part of the photon PDF.
 In this way, the ZEUS data probe only the inelastic part of the photon PDF, and therefore, we only include this inelastic contribution in comparison with the experimental data.
 In doing so we have made the approximation $f_{\gamma,\,{\rm inclusive}}(x,Q)\approx f_{\gamma,\,{\rm elastic}}(x,Q)+f_{\gamma,\,{\rm inelastic}}(x,Q)$; {\it i.e.,} the elastic and inelastic components of the photon PDF evolve separately. This approximation is good because $ f_{\gamma,\,{\rm elastic}}(x,Q)$ changes very little from $Q_0$ to $Q$ due to the rapid falloff of the proton electromagnetic form factor, while the inelastic contribution evolves additively, 
 \begin{equation}
  f_{\gamma,\,{\rm inelastic}}(x,Q)\approx f_{\gamma,\,{\rm inelastic}}(x,Q_0)+\sum_i \int_{Q_0^2}^{Q^2}\frac{dQ^2}{Q^2}\frac{\alpha}{2\pi}\,e_{i}^2\tilde{P}^{(0)}_{\gamma q}\circ 
(f_{q_i}+f_{\bar{q}_i})(x,Q)\,,
\end{equation}
up to corrections suppressed by extra factors of $\alpha$.  We have verified by explicit calculation that this additive approximation replicates the consistently evolved inclusive photon PDF, with errors that are far smaller than other theoretical uncertainties that we will discuss below.

The second additional important cut is the isolation cut on the photon, enforcing that
 90\% of the energy in the jet containing the photon belongs to the photon,
where jets are formed with the $k_T$ cluster algorithm with parameter $R=1.0$.
We will model this isolation cut in our calculation in two different ways.  First, we can model the experimental cut at the parton level, requiring $E_{\gamma}/(E_{q}+E_{\gamma})>0.9$ if the photon-quark separation satisfies $r=\Delta R_{\gamma q}=\bigl((\Delta\eta_{\gamma q})^2+(\Delta\phi_{\gamma q})^2\bigr)^{1/2}<1$.  For later reference, we call this the ``sharp'' isolation cut.   Since this does not completely remove the quark-photon collinear singularity, the theoretical calculation of the $QQ$
component will depend on the choice of the quark-to-photon fragmentation function $D_{\gamma/Q^\prime}(z,\mu_{ f})$.
For this, we use the LO fragmentation function determined by the ALEPH Collaboration~\cite{Buskulic:1995au}, parametrized by
\begin{eqnarray}
D_{\gamma/Q^\prime}(z,\mu_{ f})&=&
\frac{\alpha}{2\pi}\left[\tilde{P}_{\gamma q}(z)\ln\left(\frac{\mu_f^2}{\mu_0^2(1-z)^2}\right)+e_q^2C_0\right]\ ,\nonumber
\end{eqnarray}
where $\mu_0=0.14$ GeV and $C_0=-13.26$.  Note that this parametrization of $D_{\gamma/Q^\prime}(z,\mu_{ f})$
is an exact solution to the evolution equation at ${\cal O}(\alpha)$, so that the dependence on the fragmentation
scale $\mu_f$ cancels exactly in our calculation.  In the GGP analysis~\cite{GehrmannDe Ridder:2006wz} , other parametrizations of $D_{\gamma/Q^\prime}(z,\mu_{ f})$ with different assumptions were compared, with only a small effect on the calculated cross sections.

One of the disadvantages of having a dependence on the quark-to-photon fragmentation function, in addition to the
uncertainties due to the phenomenological fit to the function, is that it assumes that the cross section is inclusive
in the fragmentation remnant.  In our calculation, the combination of the experimental constraints on the photon and
on the lepton indirectly imposes constraints on the remnant quark in the process.  Therefore, we also consider an alternative
model of the experimental isolation cut, replacing it with a ``smooth'' isolation cut~\cite{Frixione:1998jh}, so as to avoid the
necessity of the fragmentation contribution.
The smooth isolation cut is given by requiring that the hadronic energy $E_{ h}$ inside all cones of radius $r<R$ around the photon direction satisfy
\begin{eqnarray}
E_{ h}&<& \epsilon E_{\gamma}\left(\frac{1-\cos{r}}{1-\cos{R}}\right)\ ,
\end{eqnarray}
where we take $R=1$ and $\epsilon=1/9$. These values of $R$ and $\epsilon$ are chosen to ensure that the photon will contain at least 90\% of
the energy inside a cone of $r=1.0$ centered on the photon, just as for the experimental isolation cut.  However, the smooth cut does not translate
exactly to the experimental isolation cut, because it requires the photon to carry a greater fraction of the energy as the cone size $r$ becomes smaller.  In practice, because of this, the theoretical calculation with the smooth cut is better behaved than with the sharp cut.  In addition, the smooth isolation cut is more restrictive than the sharp isolation; for a strictly positive-definite differential cross section, the smooth isolation 
prescription must always give a smaller predicted cross section.
In this way, a comparison of the two calculations can give some indication of the theoretical
uncertainty due to the isolation cut.  

Finally, we note that the jet-clustering algorithm also includes the electron, so that
the isolation cut effectively imposes $\Delta R_{\gamma e}>1$.

\begin{figure}[t]
\begin{center}
\includegraphics[width=0.47\textwidth]{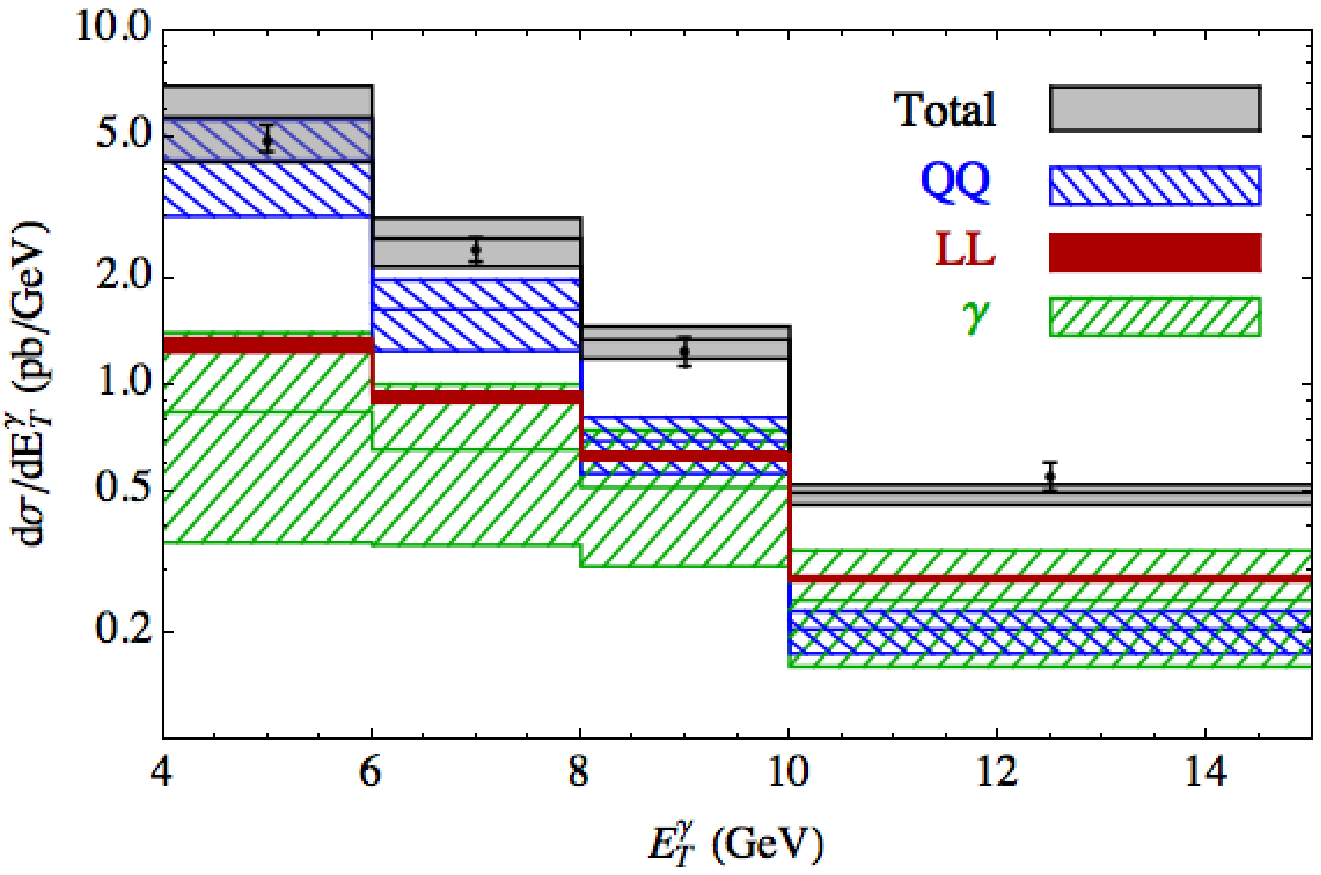}
\includegraphics[width=0.47\textwidth]{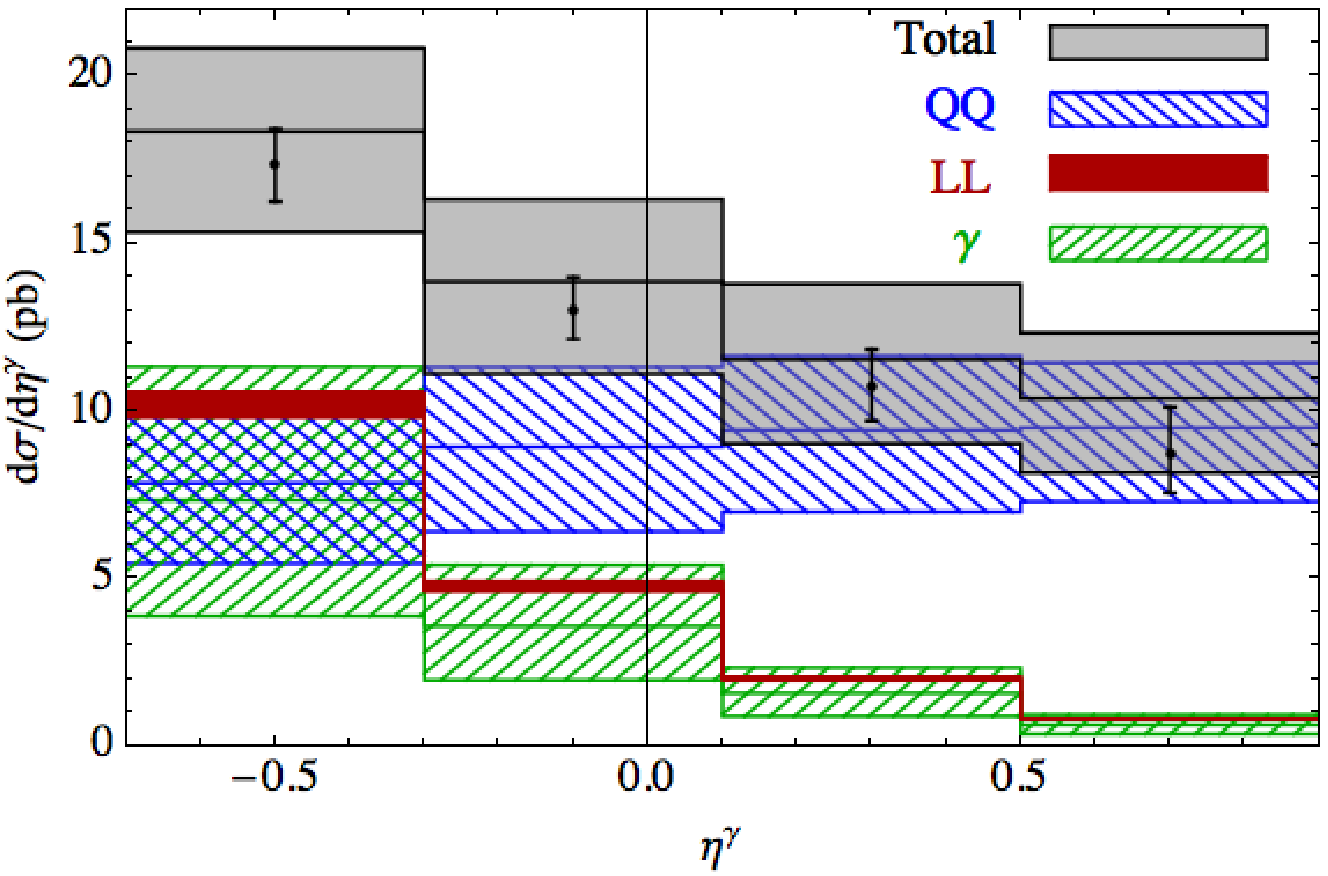}
\end{center}
\caption{Differential distributions for a zero initial inelastic photon PDF,
using the smooth isolation prescription.  The various bands display a variation in factorization scale between
$0.5 E_{\perp\gamma}\le \mu_F\le 2E_{\perp\gamma}$ and
correspond to the total prediction (light gray solid), the $QQ$ component (blue hashed), the $LL$ component (dark red solid), and the 
photon-initiated contribution only (green hashed).
 Also shown are the ZEUS data points with combined statistical and systematic errors.
\label{fig:ScaleDep}}
\end{figure}

\subsection{Comparison with the data and constraints on the photon PDF}\label{sec:ZEUSvsPhoton}

Before discussing the comparison of the theory with the data, it is useful to understand the theoretical uncertainties of the 
calculation by studying the factorization scale dependence and the dependence on the isolation prescription.  
In Fig.~\ref{fig:ScaleDep} we plot the differential cross sections for 
$d\sigma/dE_{\perp\gamma}$ and $d\sigma/d\eta_\gamma$ as functions of $E_{\perp\gamma}$ and $\eta_\gamma$,
respectively, while varying the factorization scale within $0.5 E_{\perp\gamma}\le \mu_F\le 2E_{\perp\gamma}$.
Here we have used the sharp isolation cut and calculated with zero initial inelastic photon PDF at $Q_0=1.295$ GeV.
The four bands on each of the two plots correspond to the photon-initiated contribution only (green hashed band), $LL$ component
(dark red solid band), $QQ$ component (blue hashed band), and the total calculation (light gray solid band).  The $QL$ contribution is imperceptible on
the scale of these plots.
From these plots we learn several important facts.  First, the scale dependence of the $LL$ component is reduced
dramatically compared to the photon-initiated contribution alone.  This large scale dependence of the photon-initiated
contribution cancels greatly with that of the collinear subtraction counterterm in the combined $LL$ component.
Second, the $LL$ and $QQ$ components dominate
in different regions of phase space.  For instance, the cross sections are most sensitive to the $LL$ component,
and consequently to the photon PDF, at large $E_{\perp\gamma}$ and small $\eta_\gamma$.  Thus, the shapes of these
distributions can give information about the nonperturbative contribution to the photon PDF.  Finally, we note that the
scale dependence of the $QQ$ component is still large, being only LO in $\alpha_s$, and it dominates the overall
scale uncertainty of the theoretical calculation.

\begin{figure}[t]
\begin{center}
\includegraphics[width=0.47\textwidth]{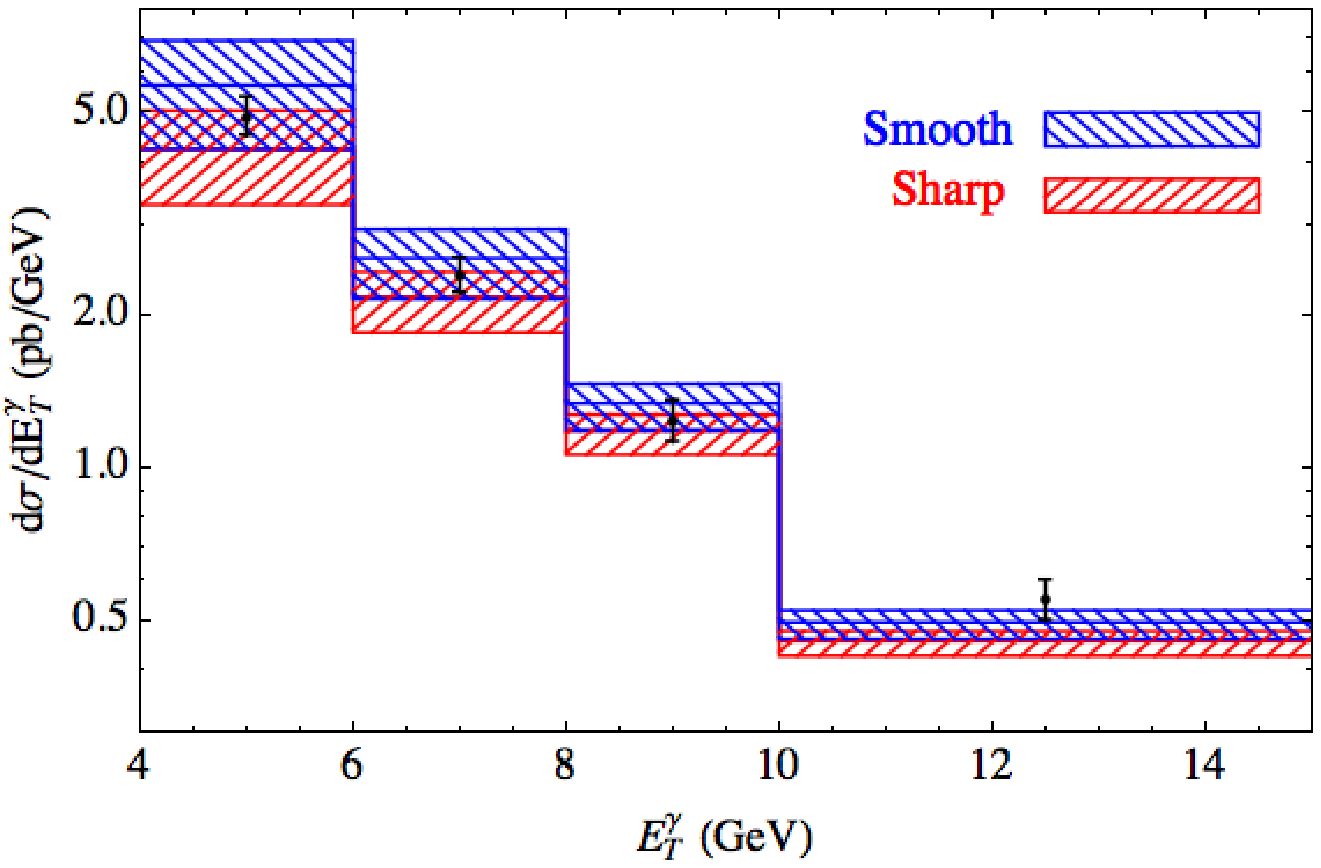}
\includegraphics[width=0.47\textwidth]{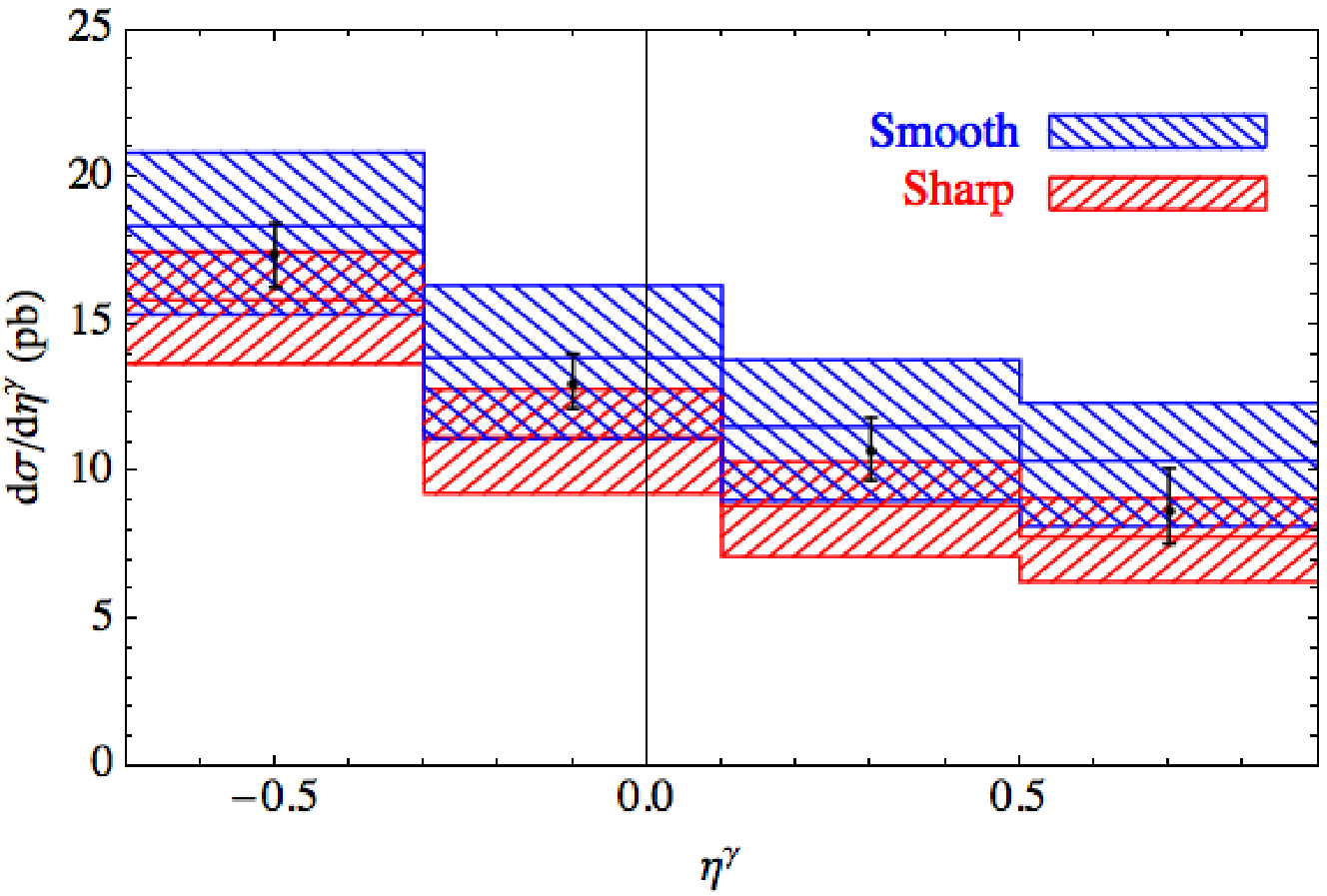}
\end{center}
\caption{Differential distributions for a zero initial inelastic photon PDF
with the factorization scale varied between $0.5 E_{\perp\gamma}\le \mu_F\le 2E_{\perp\gamma}$.
The blue hashed band is calculated using the smooth isolation prescription and the red hashed band is calculated using the sharp isolation
prescription.  Also shown are the ZEUS data points with combined statistical and systematic errors.
\label{fig:IsoDep}}
\end{figure}

In Fig.~\ref{fig:IsoDep} we plot the total predictions of the same two distributions, again for zero initial inelastic photon
PDF, but now comparing the two different isolation prescriptions in the theoretical calculation.  
In these plots we show the predictions with the smooth isolation prescription (blue hashed band) and the sharp isolation prescription
(red hashed band), again varying the factorization scale within $0.5 E_{\perp\gamma}\le \mu_F\le 2E_{\perp\gamma}$.
The first thing to note here is that the difference between the predictions is about the same size as the scale uncertainty,
with similar dependence on the kinematic variables.
Another striking feature is that the prediction using  the smooth isolation prescription is uniformly larger than 
that using the sharp isolation, in contrast
to expectations.  This is probably due to incomplete cancellations in the sharp isolation calculation between the large negative collinear 
fragmentation contribution and the positive real contribution, due to indirect constraints on the emitted final-state quark
in the real emission contribution.
Presumably higher-order QCD corrections will affect the predictions for both isolation predictions, to resolve this issue.
As noted previously, our calculation is only LO in $\alpha_s$; we expect both the factorization scale uncertainty and 
the isolation prescription discrepancy to be reduced at NLO in $\alpha_s$.
In any event, we will use the two isolation prescriptions, as well as the factorization scale dependence, as a measure of
the theoretical uncertainty of our calculation.

With this understanding of the theoretical uncertainties of the calculation we can now compare the ZEUS
data against predictions for the differential distributions, while varying the initial inelastic photon momentum fraction $p_0^\gamma$
of the photon PDF, described in Sec.~\ref{sec:photonPDF}.   For this analysis, the initial quark and gluon PDFs are just
the CT14NLO PDFs, except that the sea-quark normalizations are rescaled in order to maintain a total momentum fraction
of 1.  This rescaling has little effect in our analysis, because the photon momentum fractions considered here are small.
For instance, a photon momentum fraction of $p_0^\gamma=0.14\%$ induces a reduction of the sea-quark momentum
by only 0.9\%, while the CM photon PDF induces a reduction of the sea-quark momentum by 1.6\%.  At this stage of the analysis we have not refit the quark and gluon PDFs, since the $ep\rightarrow ep+X$ process
is dominantly sensitive to the photon PDF directly, whereas the indirect sensitivity through changes in the quark and gluon
PDFs is negligible.

\begin{figure}[t]
\begin{center}
\includegraphics[width=0.47\textwidth]{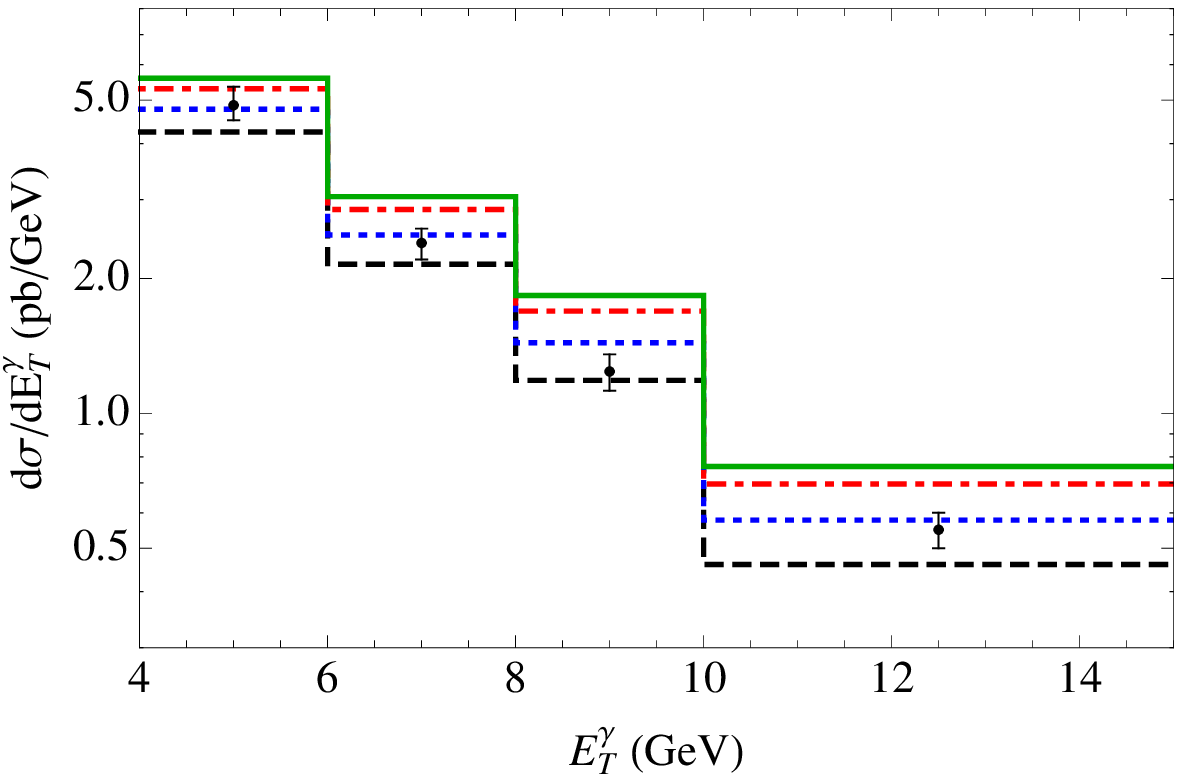}
\includegraphics[width=0.47\textwidth]{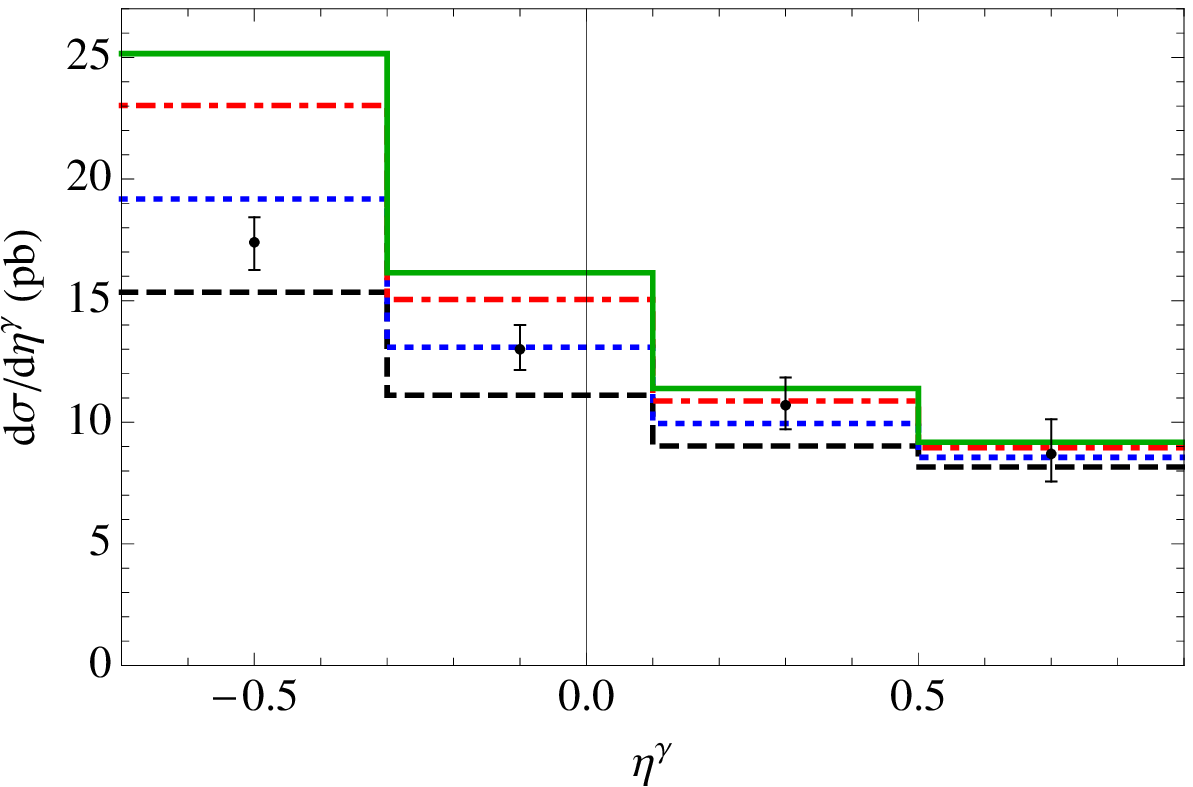}
\end{center}
\caption{Differential distributions in the photon variables, $E_{\perp\gamma}$ and $\eta_\gamma$, with
the smooth isolation prescription, with factorization scale $\mu_F=0.5E_{\perp\gamma}$.  The curves, from
bottom to top are with initial inelastic photon momentum fractions of $p_0^\gamma=0\%$ (black dashed), 0.1\% (blue dotted), 0.2\% (red dot-dashed), and for the CM photon (green solid).
 Also shown are the ZEUS data points with combined statistical and systematic errors.
\label{fig:EtEta}}
\end{figure}

In Fig.~\ref{fig:EtEta} we plot the differential cross sections for 
$d\sigma/dE_{\perp\gamma}$ and $d\sigma/d\eta_\gamma$ as a function of the
photon variables, $E_{\perp\gamma}$ and $\eta_\gamma$, using the smooth isolation prescription, with a factorization
scale of $\mu_F=0.5E_{\perp\gamma}$. The curves, from
bottom to top are with initial inelastic photon momentum fractions of $p_0^\gamma=0\%$ (black dashed), 0.1\% (blue dotted), 0.2\% (red dot-dashed), and for the CM
initial photon (green solid), which has initial momentum fraction 0.26\%.  Also shown are the ZEUS data points with combined statistical and systematic errors.  With these choices of $\mu_F$ and the isolation prescription,
we see that the theory can fit the data well for $p_0^\gamma\approx0.1$\%.  On the other hand, the theory fits poorly for the CM initial photon, overshooting the data at large $E_{\perp\gamma}$ and small $\eta_\gamma$.  Of course, the best fit for 
$p^\gamma_0$ is correlated with the choice of $\mu_F$ and the isolation prescription.  However, since these choices tend to
move the curves up or down uniformly, it is still possible to constrain the initial photon PDF by the shape of the distributions.
In particular, it is impossible to get a good fit to the prediction using the CM initial photon PDF, regardless of the choices of
$\mu_F$ and the isolation prescription.

\begin{figure}[t]
\begin{center}
\includegraphics[width=0.47\textwidth]{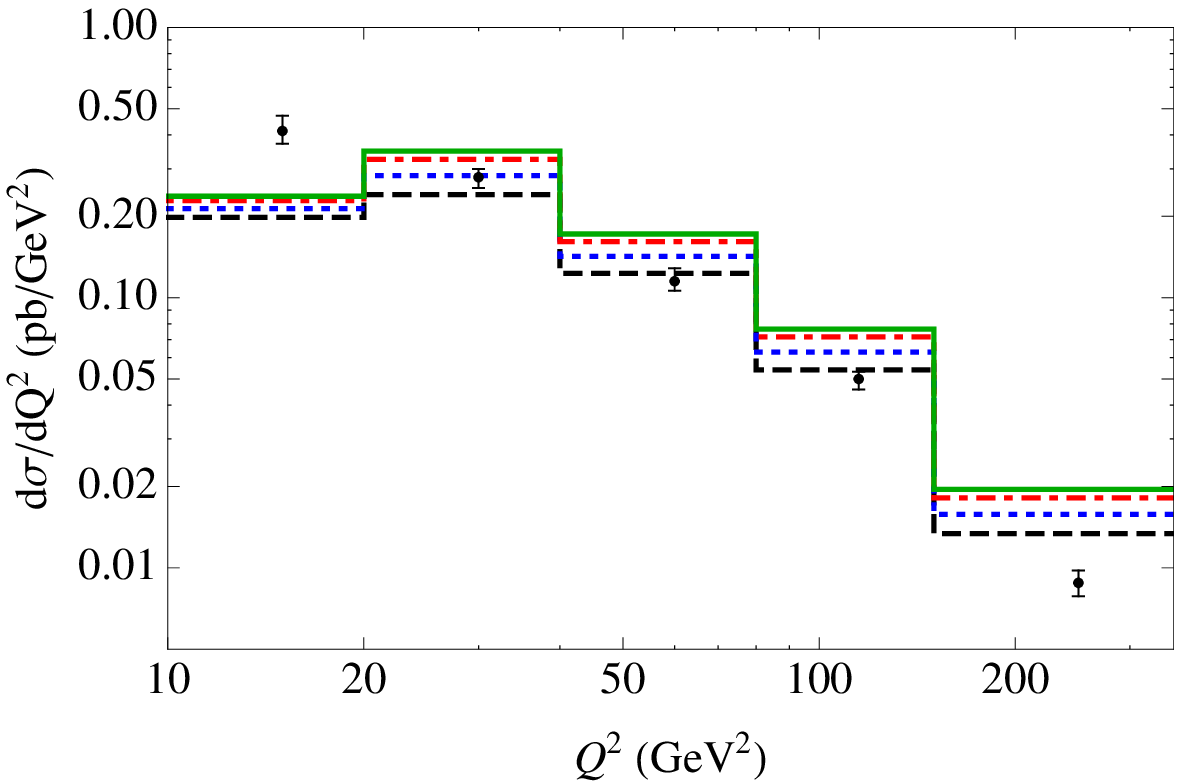}
\includegraphics[width=0.47\textwidth]{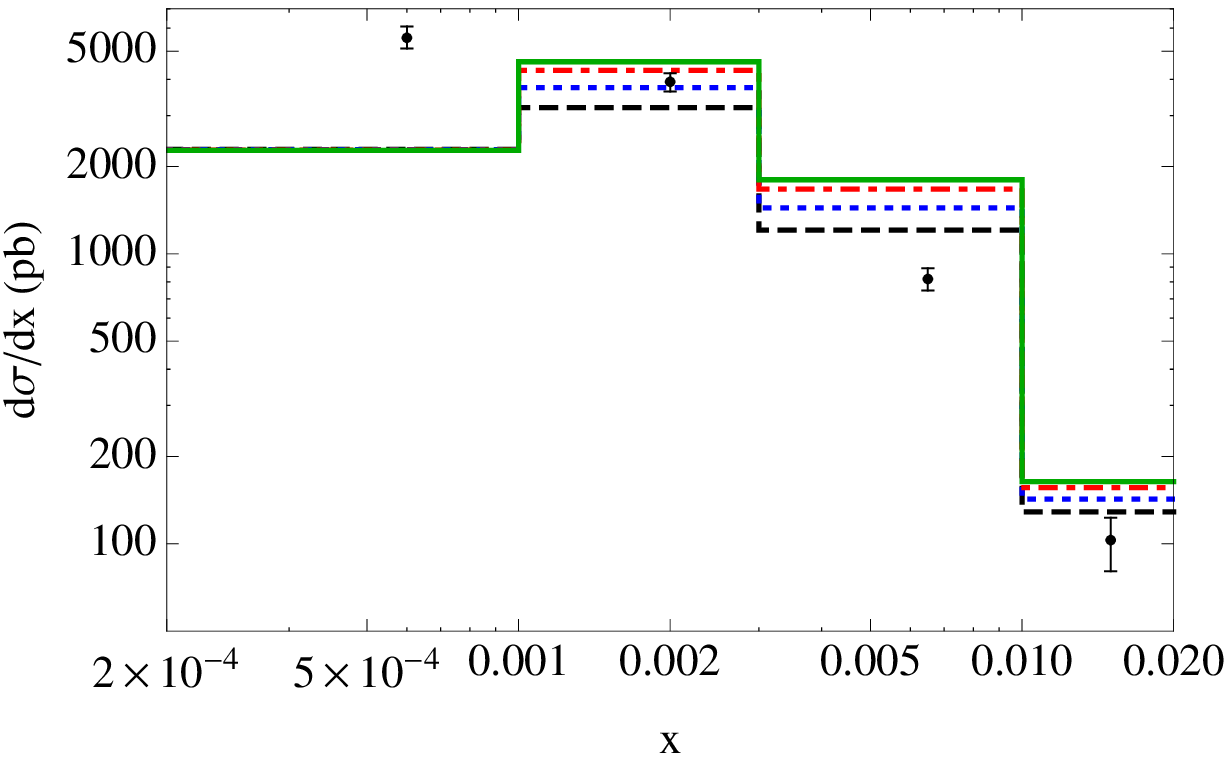}
\end{center}
\caption{Differential distributions in the lepton variables, $Q^2$ and $x$, with
the smooth isolation prescription, with factorization scale $\mu_F=0.5Q$.  The curves, from
bottom to top are with initial inelastic photon momentum fractions of $p_0^\gamma=0\%$ (black dashed), 0.1\% (blue dotted), 0.2\% (red dot-dashed), and for the CM photon (green solid).
 Also shown are the ZEUS data points with combined statistical and systematic errors.
\label{fig:Q2x}}
\end{figure}

In Fig.~\ref{fig:Q2x} we plot the differential cross sections for 
$d\sigma/dQ^2$ and $d\sigma/dx$ as a function of the
lepton variables, $Q^2$ and $x$ against the ZEUS data, using the exact same theoretical choices and initial inelastic photon PDFs as for the previous plot.  In this case we see that it is impossible to fit the data, regardless of the initial photon PDF or the
choices of factorization scale and isolation prescription.  In particular the theory fits the data very poorly at small $x$ and $Q^2$.
In fact, we note that prediction for the smallest bin in $x$ is far from the ZEUS data point, and is essentially independent of
the initial photon PDF.  We expect that the predictions in these bins are highly sensitive to higher-order QCD radiation, so that
it is difficult to fit the full lepton distributions with a fixed-order calculation.

\begin{figure}[t]
\begin{center}
\includegraphics[width=0.46\textwidth]{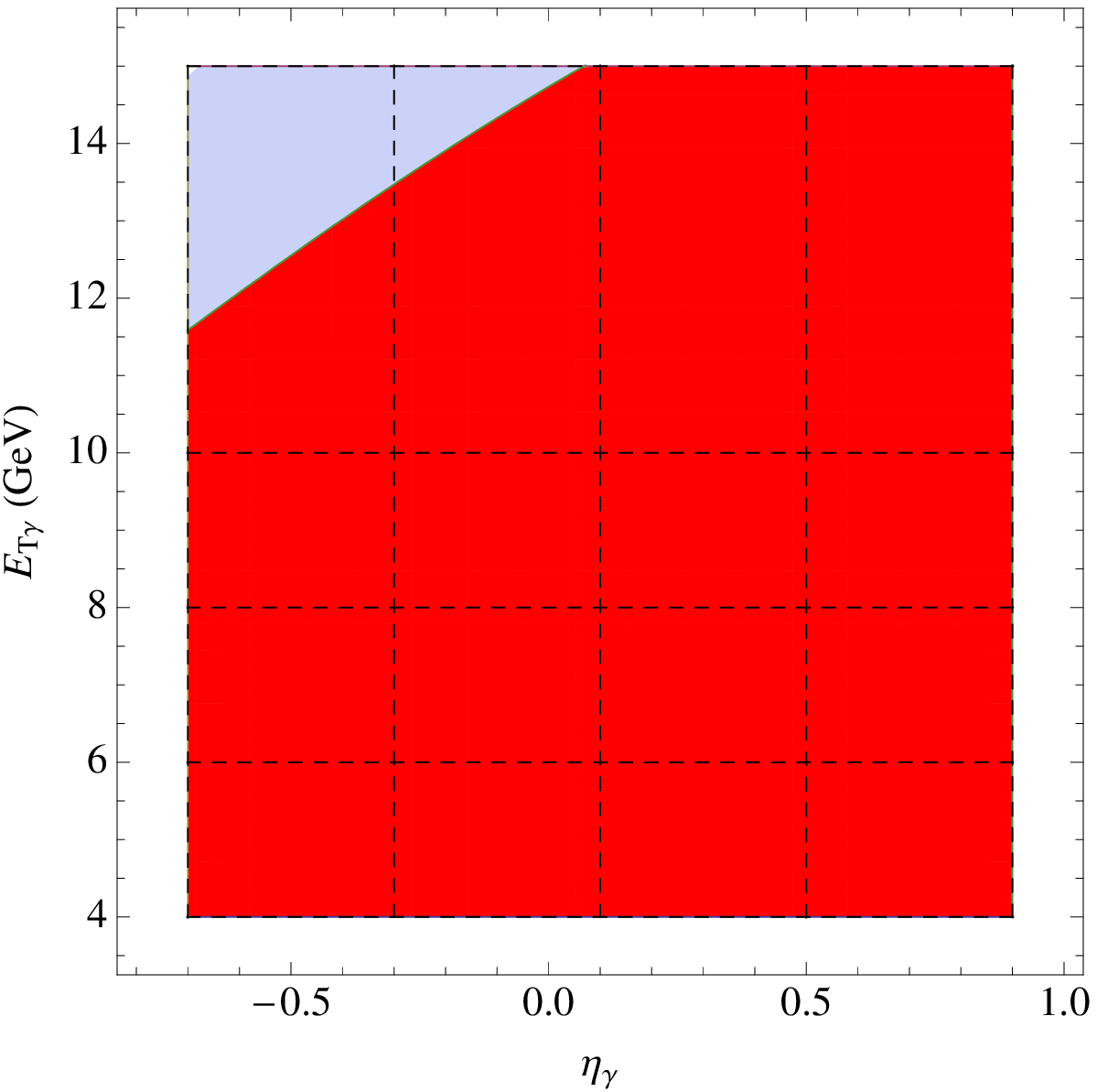}
\includegraphics[width=0.472\textwidth]{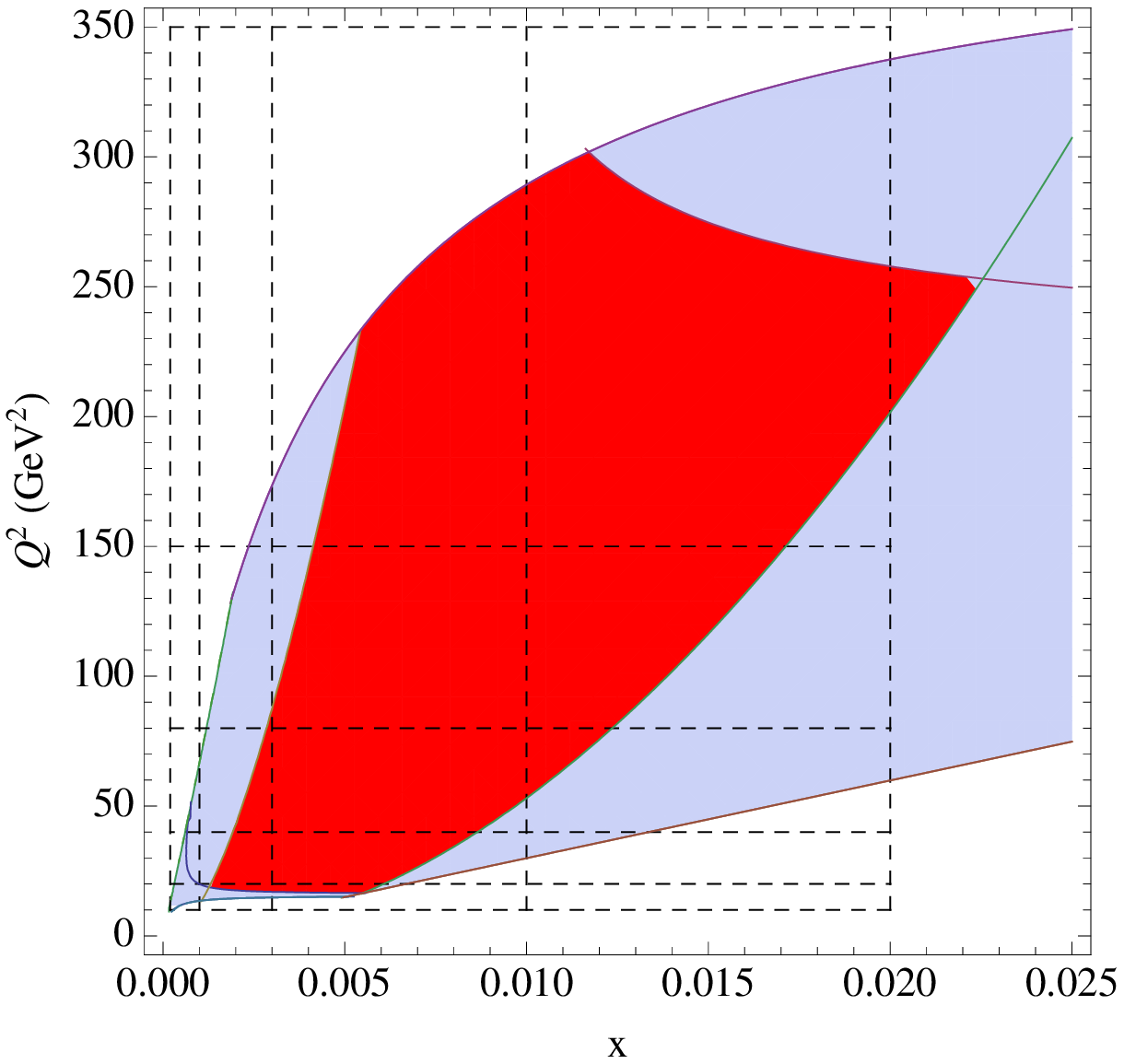}
\end{center}
\caption{Kinematic limits on $E_{\perp\gamma}$ and $\eta_\gamma$ (left) and $Q^2$ and $x$ (right).  The dashed lines indicate the bins that are plotted by the ZEUS experiment.  The dark red region only is kinematically accessible by the photon-initiated contribution, while the dark red$+$light blue regions are kinematically allowed in general.
\label{fig:Limits}}
\end{figure}

The fact that these fixed-order calculations are more reliable for the photon distributions than for the lepton distributions can
be seen further by looking at the phase-space constraints for the two sets of variables.  In Fig.~\ref{fig:Limits} we show
plots of the constraints on the photon variables $E_{\perp\gamma}$ and $\eta_\gamma$ and on the lepton variables
$Q^2$ and $x$.  In these figures, the dashed lines indicate the bins that are plotted by the ZEUS data.  The combined
dark red$+$light blue regions indicate the regions of phase space allowed by the ZEUS kinematic constraints of Eq.~(\ref{eq:zeuscuts}),
for the fully inclusive event, whereas the dark red region only is allowed for the LO photon-initiated subprocess.  
For the photon distributions, the constraints are dominated by the photon cuts on $E_{\perp\gamma}$ and $\eta_\gamma$,
with only a small cut on the photon-initiated contribution in the upper-left corner due to the requirement of $\eta^\prime_\ell<-1.0053$.  Thus, all of the bins in $E_{\perp\gamma}$ and $\eta_\gamma$ have a large photon-initiated contribution
and can be considered very inclusive.  In contrast, for the lepton distributions the additional photon constraints have a
large effect in many of the bins.  For instance, the photon-initiated contribution to the smallest $Q^2$ bin is largely removed
by the requirement of $E_{\perp\gamma}>4$ GeV, and the photon-initiated contribution to the smallest $x$ bin is
completely removed by the requirement of $\eta_\gamma>-0.7$.  These bins are dominated by events with additional
particles in order to satisfy the kinematics, so we would not expect our fixed-order calculation to do well at predicting the
lepton distributions.

\begin{figure}[t]
\begin{center}
\includegraphics[width=0.472\textwidth]{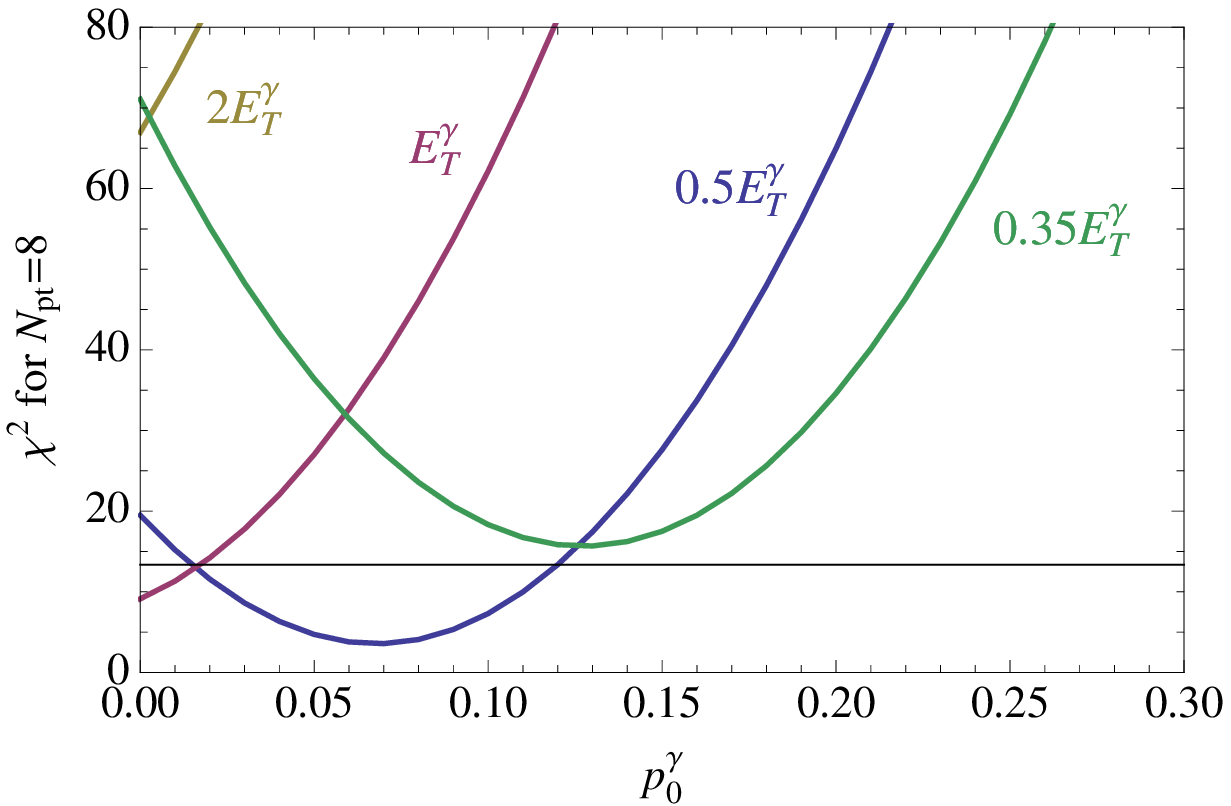}
\includegraphics[width=0.472\textwidth]{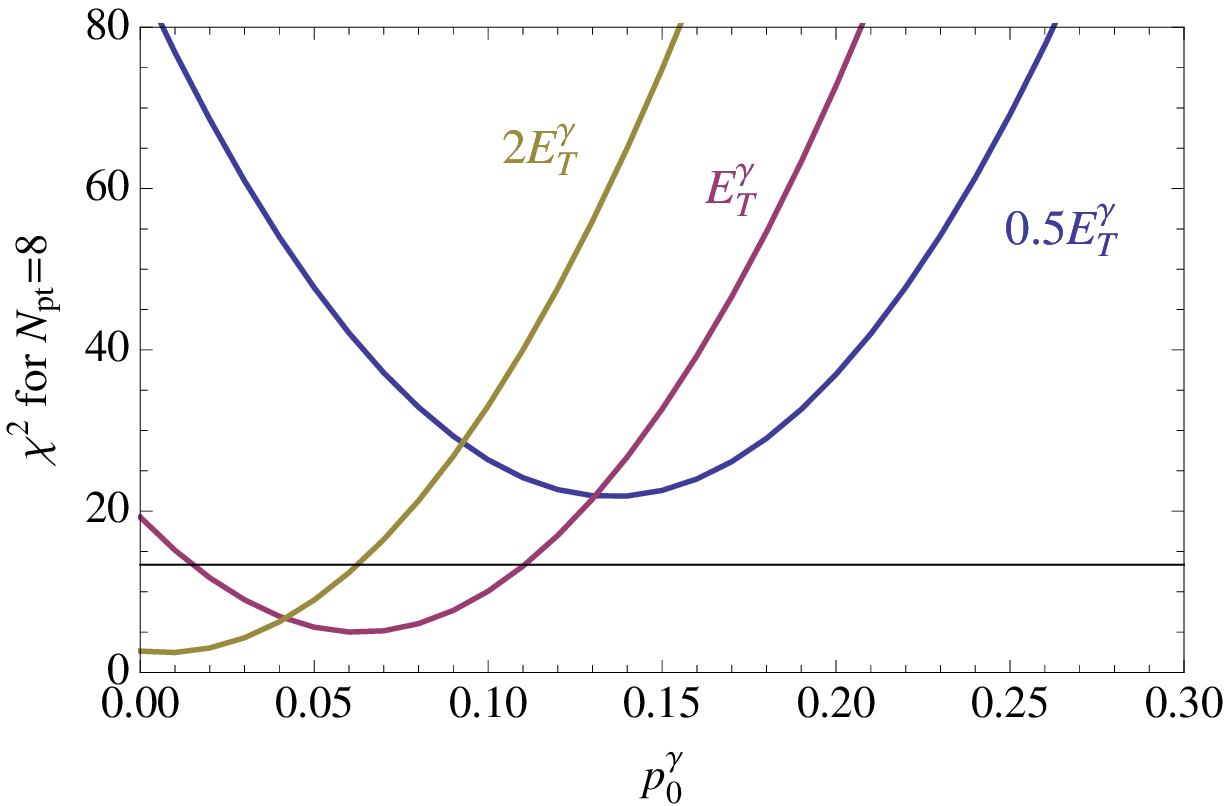}
\end{center}
\caption{Plots of $\chi^2$ versus initial inelastic photon momentum fraction $p_0^\gamma$ using the smooth isolation prescription
(left) and the sharp isolation prescription (right) for factorization scales $\mu_F=2E_{\perp\gamma}$, $E_{\perp\gamma}$, $0.5E_{\perp\gamma}$, and $0.35E_{\perp\gamma}$. The horizontal line
at $\chi^2=13.36$ is the 90\% confidence level limit for eight data points.
 \label{fig:Chi2}}
\end{figure}

Based on these arguments we will use only the distributions in the photon variables $E_{\perp\gamma}$ and $\eta_\gamma$
to constrain the initial inelastic photon PDF, for a total of eight data points.  We also reiterate that the constraints due to the remaining CT14 experimental data
set are much weaker than these data, and are neglected in the present analysis.    We define the chi-squared function for these data points by
\begin{eqnarray}
\chi^2&=&\sum_{i=1}^8\left(\frac{T_i-D_i}{\sigma_i}\right)^2\ ,
\end{eqnarray}
where $T_i$, $D_i$, and $\sigma_i$ are the theory prediction, the experimental measurement, and its combined 
statistical and systematic error for the data point $i$.  In Fig.~\ref{fig:Chi2} we plot $\chi^2$ versus the
initial inelastic photon momentum fraction $p_0^\gamma$ for both the smooth and sharp isolation prescriptions and for
several values of the factorization scale $\mu_F$.   Note that the value of $p_0^\gamma$ determined by the minimum
of $\chi^2$ depends significantly on the isolation prescription and on the factorization scale, giving best fits for the initial momentum fraction
that can vary from less than 0 to above 0.1\%.  In addition, due to this theoretical uncertainty in the current calculation, it is not possible to 
unambiguously determine an error band on $p_0^\gamma$, using the standard CT approach of applying some tolerance criterion on the rise in 
the $\chi^2$ around the best fit.  However, from Fig.~\ref{fig:Chi2} we do see that not all choices of theoretical parameters are able to fit the shape of the
data points equally well.  Therefore, we can determine a conservative limit on the value of $p_0^\gamma$ by requiring that the data and theory
not disagree beyond some level.  A $\chi^2$ distribution with eight data points will have $\chi^2<13.36$
at the 90\% confidence level.\footnote{As a comparison, the change in the total $\chi^2$ for the remaining 2947 data points used in the CT14 analysis is
$\Delta\chi^2=-2.3$ in going from $p_0^\gamma=0$\% to $p_0^\gamma=0.14$\%.}
  Therefore, we define that any theoretical prediction with $\chi^2>13.36$  is ruled out as a bad fit to the data at the 90\%
confidence level.  It is impossible to satisfy this criterion for $p_0^\gamma>0.14$\% for either choice of the isolation prescription and for any value of $\mu_F$.  Furthermore we find that the CM choice
of the photon PDF has $\chi^2>46$ for any choice of isolation and factorization scale and so is ruled out by this data.

Thus, we find our maximal initial inelastic photon PDF to have $p_0^\gamma=0.14$\% at the 90\% confidence level.  Of course, the
exact value of the momentum fraction is correlated with the shape of the initial photon PDF.   From Fig.~\ref{fig:Limits}
we see that the ZEUS DIS-plus-isolated-photon data constrains the photon PDF in the kinematic region given roughly by
$10^{-3}< x<2\cdot10^{-2}$ for $16< Q^2 <300$ GeV$^2$.  Outside of this region the photon PDF is very weakly constrained, but
we believe that the radiative ansatz gives a reasonable expectation for its overall shape.
As for the minimal possible value of the initial inelastic photon momentum fraction, it could, in principle, be negative, which is not ruled out by the analysis of this section.  For instance, one could begin the evolution with zero initial photon PDF at a lower value of the scale $Q_0$.  However,
we take the reasonable assumption that it should be nonzero at the low scale of $Q_0=1.295$ GeV.  We thus propose the initial PDFs
with $p_0^\gamma=0$\% and $p_0^\gamma=0.14$\% as our 90\% C.L. photon PDFs.  A similar analysis gives $p_0^\gamma\le0.11$\% at the 68\% confidence level,
but the data are still consistent with $p_0^\gamma=0$\% at the 68\% C.L.

\begin{figure}[t]
\begin{center}
\includegraphics[width=0.46\textwidth]{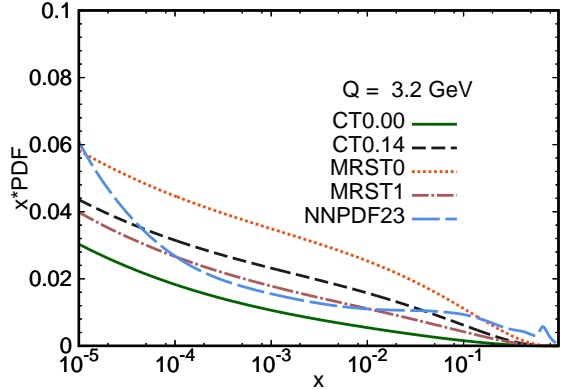}
\end{center}
\caption{Comparison of various NLO photon PDFs at the scale $Q=3.2$ GeV:  CT14QED with $p_0^\gamma=0\%$ (solid green),
CT14QED with $p_0^\gamma=0.14\%$ (short-dashed black), MRST2004QED0 using current quark masses (dotted orange), MRST2004QED1 using
constituent quark masses (dot-dashed brown), and NNPDF2.3QED with $\alpha_s=0.118$ and average photon (long-dashed blue).
 \label{fig:comp1}}
\end{figure}

In Fig.~\ref{fig:comp1} we compare, at the scale $Q=3.2$ GeV, the CT14QED photon PDFs with $p_0^\gamma=0$\% and $p_0^\gamma=0.14$\% against the NLO MRST2004QED photon PDFs, using the current quark masses (labeled MRST0) and using the constituent quark masses (labeled MRST1), and against the NLO NNPDF2.3QED average photon PDF with $\alpha_s=0.118$.  
We should emphasize that the CT14QED photon PDFs only contain the inelastic contribution in these plots.
The NNPDF2.3 average photon PDF has more
structure in its shape at large and small values of $x$ than do the other PDFs, but it is still consistent with the ZEUS data in the $x$ range that is probed by the experiment.

\begin{figure}[t]
\begin{center}
\includegraphics[width=0.46\textwidth]{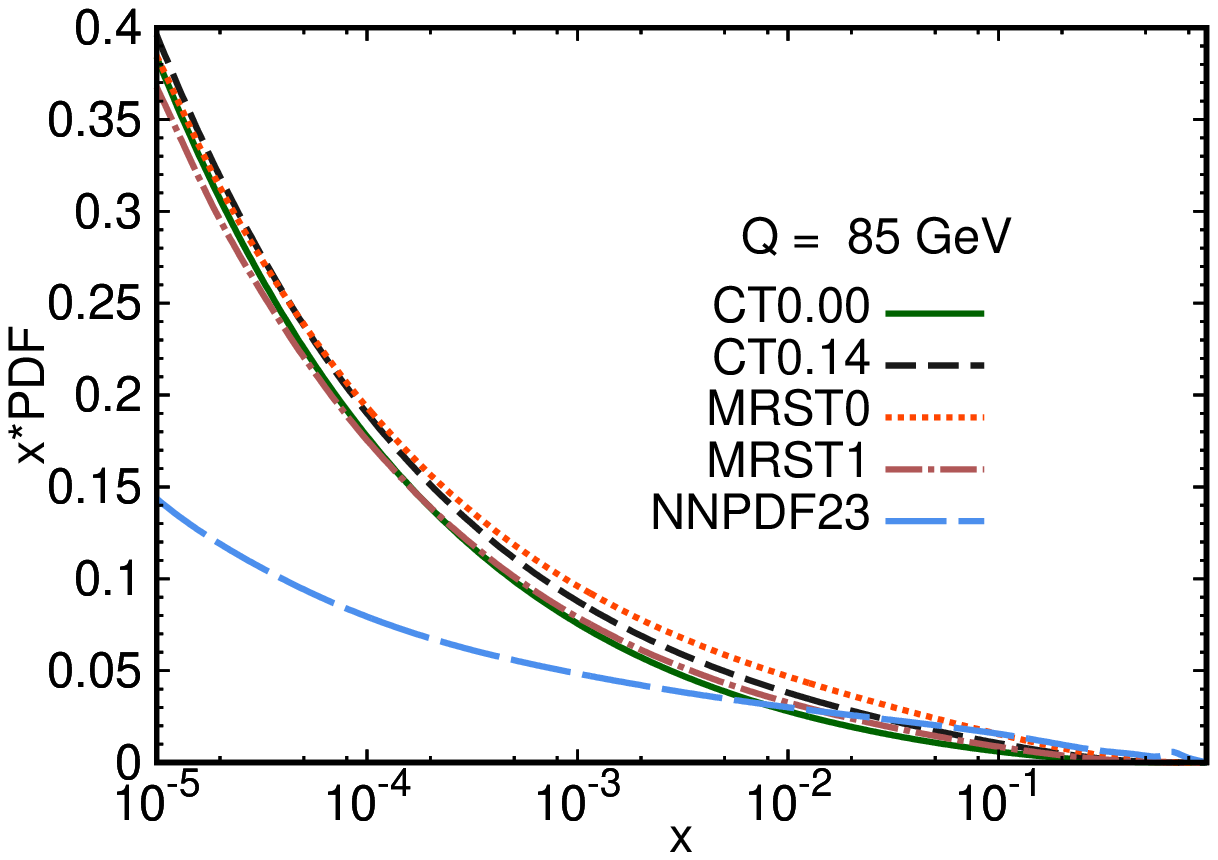}
\includegraphics[width=0.46\textwidth]{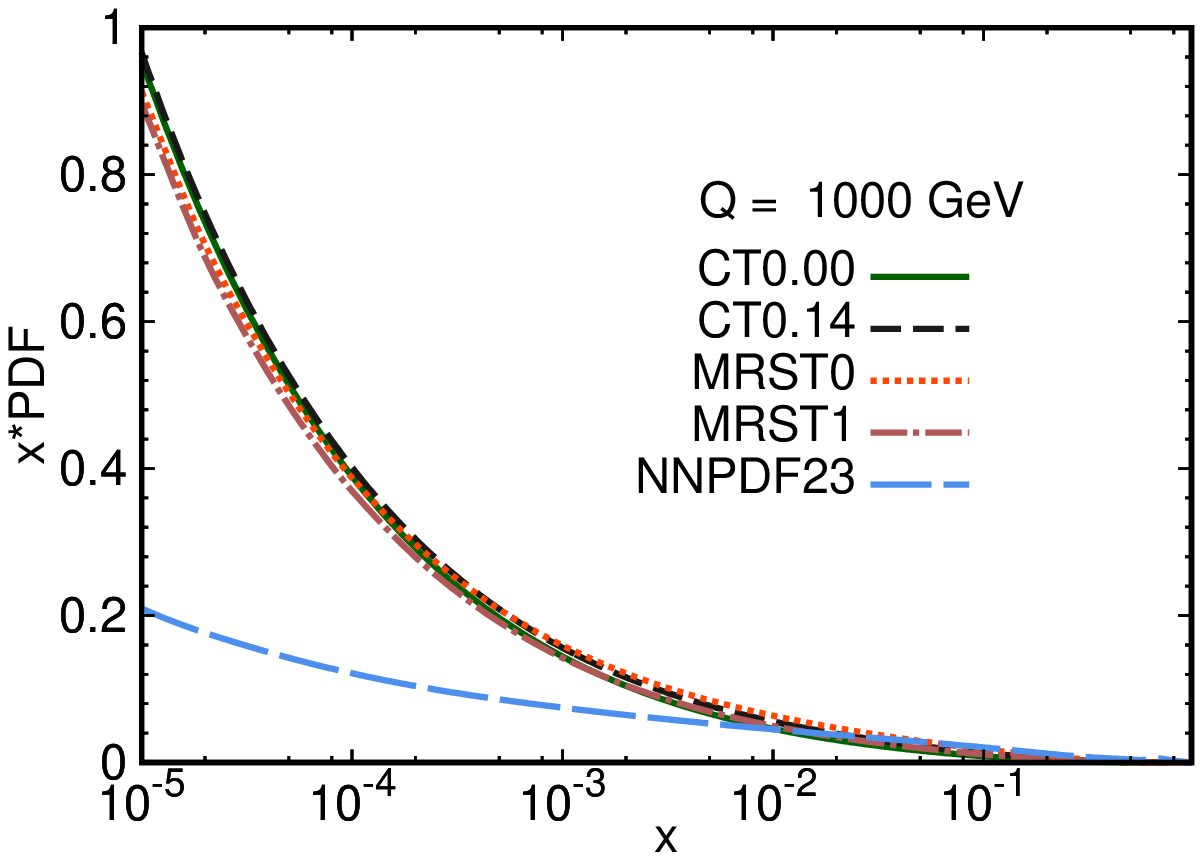}
\end{center}
\caption{Comparison of various NLO photon PDFs at the scales $Q=85$ GeV (left) and $Q=1$ TeV (right): 
CT14QED with $p_0^\gamma=0\%$ (solid green),
CT14QED with $p_0^\gamma=0.14\%$ (short-dashed black), MRST2004QED0 using current quark masses (dotted orange), MRST2004QED1 using
constituent quark masses (dot-dashed brown), and NNPDF2.3QED with $\alpha_s=0.118$ and average photon (long-dashed blue).
 \label{fig:comp2}}
\end{figure}

In Fig.~\ref{fig:comp2} we compare the same set of photon PDFs at the higher scale of 85 GeV and the very high scale of 1 TeV.
Here we can make some very interesting observations.  The most obvious thing in these figures is that the CT and MRST photon
PDFs become very similar at large $Q^2$, whereas the NNPDF photon PDF is distinctly different and much smaller at small values of $x$.
This difference is due to the different approaches to the evolution of the PDFs taken by the different groups.  Whereas
in the MRST and CTEQ-TEA approaches, the QCD and QED scales are chosen to be identical and evolved together, in the NNPDF2.3QED PDFs
that are included in the LHAPDF library~\cite{LHAPDF6}, the QCD and QED scales are separate and the two scales are evolved successively; first the QED scale 
is evolved from $Q_0$ to $Q$ and then the QCD scale is evolved from $Q_0$ to $Q$.  As discussed in Ref.~\cite{Bertone:2013vaa}, the successive evolution
of QED and QCD differs from the combined evolution by terms that are subleading by ${\cal O}(\alpha\alpha_s)$ and can induce large unresummed
logarithms between the two scales.   This difference in the evolution
at small $x$ is also seen to be consistent with the behavior seen in the right panel of Fig.~2 in Ref.~\cite{Ball:2013hta}, where the NNPDF photon PDF also
is smaller at small $x$ and large $Q^2$ than when it is evolved using the code {\tt partonevolution}~\cite{Weinzierl:2002mv,Roth:2004ti}.
We expect that the difference between the NNPDF2.3QED photon PDF and the other photon PDFs at high $Q$ would be less significant if the 
NNPDF2.3QED PDFs were evolved from the low scale simultaneously in QED and QCD.

Another observation from Fig.~\ref{fig:comp2}, concerning the CT14QED and MRST2004QED photon PDFs is that the impact of the
initial photon distribution becomes less significant as $Q^2$ increases and more photons are produced through radiation off the quarks.
From these plots we see that the fractional deviation between the different photon PDFs decreases with increasing $Q^2$.
In fact at very small $x$ and large $Q^2$ the differences in the sea-quark distributions of the PDFs presumably have more impact on the photon
PDF than does the initial photon distribution.

\section{Discussion and Conclusion}\label{sec:conclude}

In this paper, we have presented CT14QED, which is the first
set of CT14 parton distribution functions obtained by including QED evolution at leading order (LO) with next-to-leading-order (NLO) QCD evolution in the global analysis by the CTEQ-TEA
group. This development will provide better theory predictions to compare with the precision data, such as Drell-Yan pair production, measured at the LHC.  The CT14QED PDFs are based on the CT14 NLO initial distributions with the addition of an initial photon PDF.  (There is also an inconsequential rescaling of the quark sea PDFs, in order to maintain the momentum sum rule.)  The inelastic contribution to the photon PDF is parametrized 
at the initial scale $Q_0$ using a generalization of the radiative ansatz introduced by the MRST group in their previous study.  The initial photon PDF then depends on two independent parameters [cf. Eq.~(\ref{eq:photonPDFintro})], which are related to the scales at which the radiation off the up and down valence quarks is cut off.  However, given the weak constraints on the initial photon PDF we find it convenient at this time to 
set the scales equal, so that the initial photon PDF is parametrized by a single parameter, which we take to be the momentum fraction
carried by the inelastic photon at the initial scale $Q_0$. For comparison purposes we have also defined a ``Current Mass'' (CM) photon PDF, comparable to the MRST current mass PDF, for which the initial photon momentum fraction is $p^0_\gamma=0.26\%$.

A set of neutron PDFs can also be obtained with a small amount of isospin breaking, suggested by the radiative ansatz applied to first order in
$\alpha$, and which automatically ensures that the number and momentum sum rules are satisfied.  However, as previously seen by both the
MRST and NNPDF groups, we find that the constraints from isospin violation effects (generally small and most important at large $x$)
in nuclear scattering and from the requirement of the momentum sum rule, imposed by the DIS and Tevatron data in the CT14, are relatively
weak.  

Thus, in order to constrain the photon PDF, we focused on the scattering process $ep\rightarrow e\gamma X$, which was measured by the ZEUS experiment at HERA.  This process is dominantly sensitive to the inelastic photon PDF directly, with negligible indirect sensitivity through the modification 
of the
quark and gluon PDFs by QED effects.  It also has the advantage that the initial-state photon subprocess contribution occurs at leading order, so
that it does not compete with other much larger contributions.  In this paper we have produced for the first time a consistent  and systematic calculation for this cross section that combines both the photon- and quark-initiated subprocesses, and simultaneously reduces the factorization
scale dependence of either calculation.  Details of this calculation were presented in Sec.~\ref{sec:theory}.  The photon isolation cut, which
required that the final-state photon must contain at least 90\% of the energy in the jet to which it belongs (where jets are formed with the
$k_T$ cluster algorithm with parameter $R=1.0$), was modeled using two different models of photon isolation.  We used the two different isolation
prescriptions, as well as the factorization scale dependence as a measure of the theoretical uncertainty of our calculation.  

By comparing the ZEUS data for the distributions of transverse energy and pseudorapidity of the final-state photon against our calculation
of the differential distributions, we were able to constrain the initial inelastic photon momentum fraction inside the proton to be $p^0_\gamma < 0.14\%$
at the 90\% confidence level.  Hence, the CM choice of photon PDF has been ruled out by this data.  For completeness,
we also compared the CT14QED PDFs to some of the NLO (in $\alpha_s$) photon PDFs published by the MRST and the NNPDF groups.
Phenomenological applications of the CT14QED PDFs will be discussed in future publications.

As shown in Fig.~\ref{fig:IsoDep}, the theoretical uncertainties due to the factorization scale dependence and the isolation prescription are currently larger than
the experimental uncertainties of the Zeus data.  Thus, extending our calculation to NLO in $\alpha_s$ should be able to further constrain the initial photon PDF.
This is a project that we are currently undertaking.

Parametrizations for the (inelastic photon) CT14QED PDF sets (both proton and neutron versions)
will be distributed in a standalone
form via the CTEQ-TEA Web site \cite{CT14website}, or as a part of
the LHAPDF6 library \cite{LHAPDF6}. For backward compatibility with
version 5.9.X of LHAPDF, our Web site also provides CT14 grids in the
LHAPDF5 format, as well as an update for the CTEQ-TEA module
of the LHAPDF5 library, which must be included during compilation
to support calls of all eigenvector sets included with CT14 \cite{LHAPDF5}.
We will also distribute sets with the inclusive photon PDFs, CT14QEDinc.
For the proton, CT14QEDinc at the initial scale $Q_0$ is the sum of 
the (inelastic) CT14QED and the elastic component of the photon PDF,
given by the Equivalent Photon Approximation. The proton CT14QEDinc
PDFs are then evolved from $Q_0$ to $Q$ as discussed in Sec.~\ref{sec:QCDplusQED}.
For the neutron, CT14QEDinc
is equal to CT14QED, since the neutron has zero electric charge, and therefore
it has no elastic component of the photon PDF.

\begin{acknowledgments}

This work was supported by the National
Science Foundation under Grant No. PHY-1417326.
We also thank Sayipjamal Dulat, Tie-Jiun Hou, Joey Huston, Pavel Nadolsky
and Jianwei Qiu for helpful discussions.

\end{acknowledgments}

\clearpage

\section*{Appendix: Comparison of our QCD plus QED evolution code with other codes
\label{sec:app1}}

\begin{figure}[t]
\begin{center}
\includegraphics[width=0.6\textwidth]{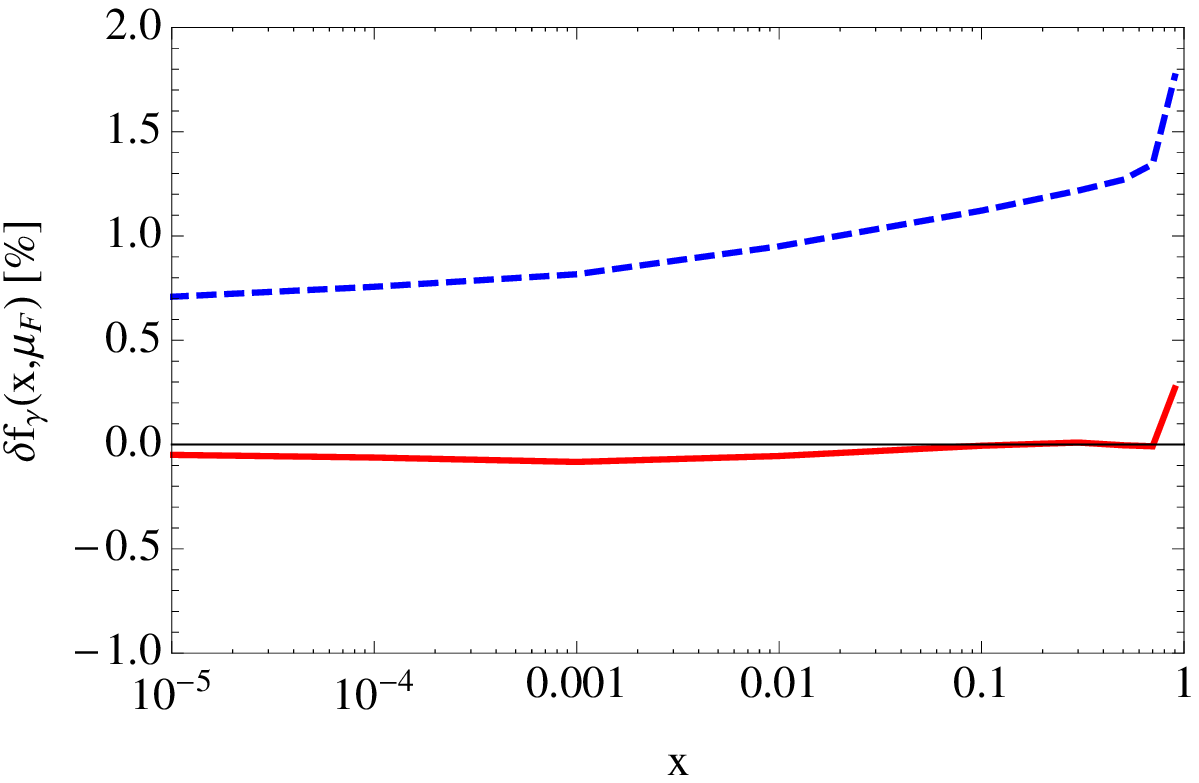}
\end{center}
\caption{Percent difference in our prediction for the photon PDF, relative to the {\tt partonevolution} prediction, for the toy model parametrizations evolved from  $Q_0=1.3$ GeV to $\mu_F=100$ GeV, as discussed in the text.
 The solid red curve uses the {\tt partonevolution} calculation with lepton PDFs removed from the evolution, while
 the dashed blue curve includes the lepton PDFs in {\tt partonevolution}.
 \label{fig:PEcomp}}
%
\begin{center}
\includegraphics[width=0.6\textwidth]{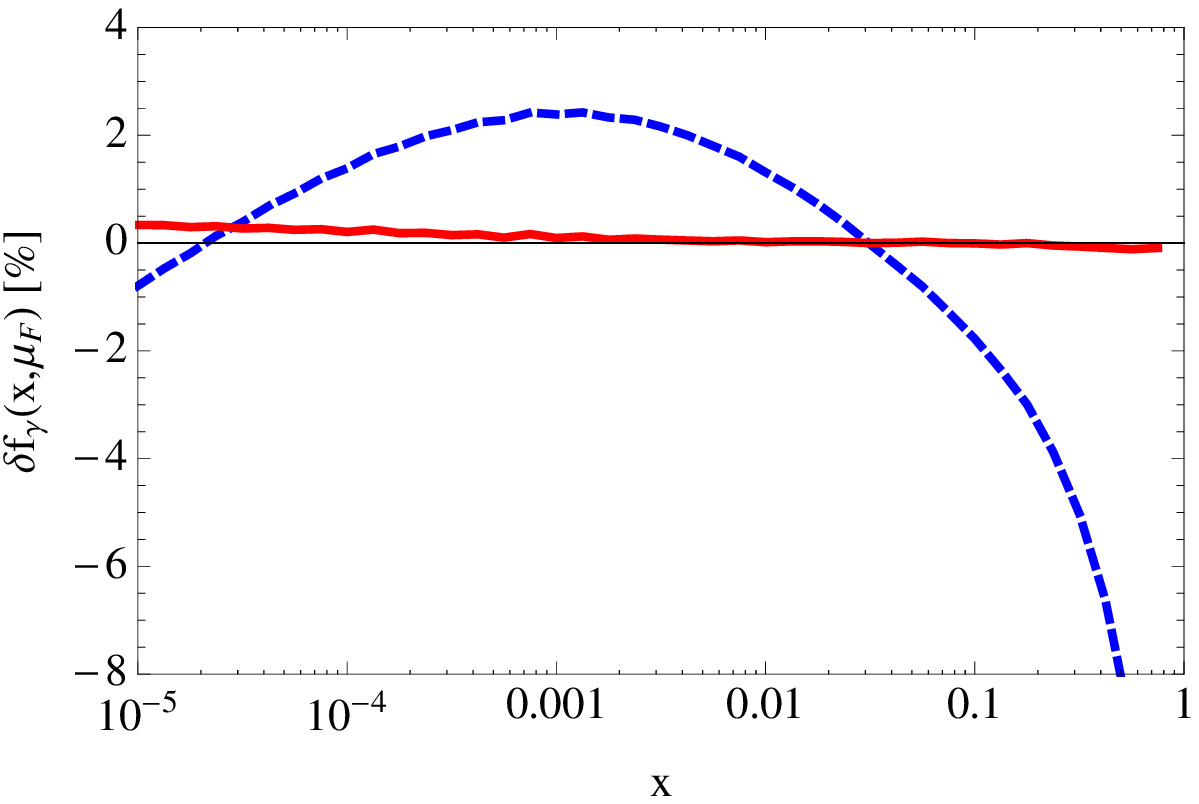}
\end{center}
\caption{Percent difference in our prediction for the photon PDF, relative to the prediction by {\tt APFEL} with the ``{\tt QUniD}'' setting (red solid) and ``{\tt QavDP}'' setting (blue dashed).  The PDFs are CT10NLO with zero initial photon PDF,
 evolved from $Q_0=1.3$ GeV to $\mu_F=100$ GeV.
\label{fig:APFELcomp}}
\end{figure}

 We have checked our code against the public evolution code {\tt partonevolution}~\cite{Weinzierl:2002mv,Roth:2004ti}, which also solves the evolution equations at LO in QED and NLO in QCD.  The main difference (other than technicalities of implementation) in the  {\tt partonevolution} code is that it also includes the charged leptons as partons in the proton.  Using the toy model of Ref.~\cite{Blumlein:1996rp} with zero initial photon PDF, ensuring all
input parameters agree, using the same formulation for the running of $\alpha_s$ and $\alpha$, and removing the lepton
PDFs and their contribution to $\tilde{P}^{(0)}_{\gamma \gamma}$ from {\tt partonevolution}, we find excellent agreement 
between the two programs.  Evolving from $Q_0=1.3$ GeV to $\mu_F=100$ GeV, we find differences of less than
$0.2$\% over most of the range of $x$ for all of the PDFs, including the photon.  Reinstating the lepton PDF contribution to the
evolution equations in {\tt partonevolution}, which in principle should be included for consistency, we find their effects on the quark and gluon PDFs to be negligible, with changes of less than $10^{-3}$\%.  The only noticeable effect is
the reduction of the photon PDF by about $1$\% with mild $x$ dependence, due to splitting of the photons into 
lepton-antilepton pairs.  This is presumably
much less than other uncertainties in our analysis, so it is reasonable to leave out the lepton PDF contribution in our code.
The percent difference in our prediction for the photon PDF, relative to the {\tt partonevolution} prediction with and without the inclusion of the
lepton PDFs, is shown in Fig.~\ref{fig:PEcomp}.

We have also checked our code against the program {\tt APFEL}~\cite{Bertone:2013vaa}, which includes QED at LO
and QCD at up to NNLO.  The main difference in the {\tt APFEL} program is that the QCD and QED factorization scales can be taken to
be independent, and the evolution with respect to each scale can be done successively.  However, using the setting ``{\tt QavDP}'' in {\tt APFEL} averages the two possible orderings for performing the evolutions, which should agree with our approach to ${\cal O}(\alpha^2)$.  In addition, the {\tt APFEL} code has
been recently updated to allow the simultaneous evolution of the QED and QCD scales, using the ``{\tt QUniD}'' setting.
We have compared our code with {\tt APFEL} 2.6.0, starting with the CT10NLO PDFs~\cite{ct10nn} with zero initial photon PDF at $Q_0=1.3$ GeV and 
evolving in QED at LO and QCD at NLO to $\mu_F=100$ GeV.  We have done the comparison using both the ``{\tt QavDP}'' and the ``{\tt QUniD}'' settings 
for {\tt APFEL}.  We obtain excellent agreement for the quark and gluon PDFs, with differences of less than $0.2$\% over most of the range of $x$ for both 
{\tt APFEL} settings.  In Fig.~\ref{fig:APFELcomp} we show the results
for the photon PDF.  We obtain pretty good agreement with  {\tt APFEL} with the ``{\tt QavDP}'' setting, with differences of less than $2.5$\% except at large
$x>0.1$. This is consistent with the ${\cal O}(\alpha^2)$ differences expected in the different evolution procedures.
We obtain excellent agreement with  {\tt APFEL} with the ``{\tt QUniD}'' setting, with differences of less than $0.34$\% over the full range of
$x>10^{-5}$ shown.  We note that in Fig.~\ref{fig:APFELcomp} we replace the evolution subroutine for $\alpha$ in the {\tt APFEL} program with the code used in
the CT global analysis code; however, using the original $\alpha$ subroutine in {\tt APFEL} still gives differences of less than 1\% for the evolved photon PDF over the full range of $x$ when using the ``{\tt QUniD}'' setting.  This is certainly much smaller than the uncertainties in the initial photon PDF itself.

We have not checked our code directly against the MRST evolution code or the recently developed {\tt QCDNUM+QED} evolution code~\cite{Sadykov:2014aua}, but we do note that the comparison between {\tt QCDNUM+QED} and {\tt APFEL} ``{\tt QavDP}'' for the evolution of the photon PDF in Ref.~\cite{Sadykov:2014aua} looks qualitatively similar to the results that we have found in Fig.~\ref{fig:APFELcomp}.  In addition benchmarking studies between {\tt APFEL}
and these two evolution codes in Ref.~\cite{Carrazza:2015dea} show agreement at a similar level to that which we have found with our code here.

\end{document}